\documentclass[a4paper,fleqn]{cas-sc}
\usepackage{setspace}

\usepackage[numbers,compress]{natbib}
\usepackage{placeins}

\def\tsc#1{\csdef{#1}{\textsc{\lowercase{#1}}\xspace}}
\tsc{WGM}
\tsc{QE}
\tsc{EP}
\tsc{PMS}
\tsc{BEC}
\tsc{DE}

\begin{document}

\let\WriteBookmarks\relax
\def\floatpagepagefraction{1}
\def\textpagefraction{.001}
\shortauthors{B. C. P. Sturmberg et~al.}
\shorttitle{A mutually beneficial approach to electricity network pricing}
\title [mode = title]{A mutually beneficial approach to electricity network pricing in the presence of large amounts of solar power and community-scale energy storage}    



\author[1]{B. C. P. Sturmberg}[type=editor,
                        orcid=0000-0001-9142-2855]
\cormark[1]
\ead{bjorn.sturmberg@anu.edu.au}


\address[1]{Battery Storage and Grid Integration Program; School of Engineering; College of Engineering and Computer Science; The Australian National University}

\author[1]{M. E. Shaw}
\ead{marnie.shaw@anu.edu.au}

\author[1]{C. P. Mediwaththe}
\ead{chathurika.mediwaththe@anu.edu.au}

\author[1]{H. Ransan-Cooper}
\ead{hedda.ransan-cooper@anu.edu.au}

\author[1]{B. Weise}
\ead{benjamin.weise@anu.edu.au}

\author[1]{M. Thomas}
\ead{michael.thomas@anu.edu.au}

\author%
[1]{L. Blackhall}
\ead{lachlan.blackhall@anu.edu.au}


\begin{abstract}
Electricity distribution networks that contain large photovoltaic solar systems can experience power flows between customers. These may create both technical and socio-economic challenges. This paper establishes how these challenges can be addressed through the combined deployment of Community-scale Energy Storage (CES) and local network tariffs. Our study simulates the operation of a CES under a range of local network tariff models, using current Australian electricity prices and current network prices as a reference. We assess the financial outcomes for solar and non-solar owning customers and the distribution network operator. We find that tariff settings exist that create mutual benefits for all stakeholders. Such tariffs all apply a discount of greater than 50\% to energy flows within the local network, relative to regular distribution network tariffs. The policy implication of these findings is that the, historically contentious, issue of network tariff reform in the presence of local solar power generation can be resolved with a mutually beneficial arrangement of local network tariffs and CES. Furthermore, the challenge of setting appropriate tariffs is eased through clear and intuitive conditions on local network tariff pricing.
\end{abstract}






\maketitle
\doublespacing
\section{Introduction}
The deployment of small-scale electricity generation and storage assets, such as rooftop solar photovoltaic systems and home batteries, commonly referred to as Distributed Energy Resources (DERs), into distribution networks creates the potential for DER-owning customers to export power into the network and for this to flow a short distance to other customers. 
The prevalence of such two-way power flows has increased significantly as the quantity and power capacity of DER systems has increased.
In Australia, for example, one in four households now has a photovoltaic solar system \cite{CER2021}.

This arrangement is beneficial in reducing distribution losses but also raises economic and technical challenges.
The economic issues center on how to price the transportation of energy flows that are contained within a small network segment (e.g., a neighbourhood). This issue is controversial because, on the one hand, the flows utilise only a small portion of the network and therefore ought to incur minimal charges, while on the other hand, they are as critically dependent upon the provision of the network segment as all other flows. Additionally, it may be argued that DER owners ought to be charged network fees for energy that they export \cite{AEMC2021}, as the profits derived from exports rely on the services of the network. The issue is particularly contentious because of the equity implications of potentially differentiating between customers who own DER and those who do not own DER \cite{ZANDER2020111508}.
The injection of significant amounts of power from DER installed `behind-the-meter' (BTM) on customers' properties' also raises technical issues including breaches in statutory voltage limits, phase imbalance and reverse power flows through transformers \cite{Cipcigan}. These may require upgrades in order for the network to host higher levels of DER, thereby driving up network costs for all customers.

Proposals to address the economic issues have generally either applied a discount to network tariffs for local energy flows \cite{RUTOVITZ2018324} or created a secondary energy market for peer-to-peer transactions. These have yet to see widespread adoption because the former may introduce a zero-sum wealth transfer between networks and customers, while the latter faces implementation and regulatory complexities.
The rise of DERs has also spurred the emergence of a new class of energy storage assets, called ``Community Energy Storage'' (CES) \cite{barbour_community_2018} (or alternatively ``Neighbourhood-scale Batteries'' \cite{Ransan-Cooper2021}) that connect directly into the distribution network.
The location of these assets, close to customers, makes them well suited to managing network conditions, such as breaches in network voltages due to high penetration of rooftop photovoltaics \cite{mediwaththe_network-aware_2021}, and to providing economic benefits to communities \cite{SCHELLER, dong_impact_2020}. For these reasons, CES systems are gaining momentum, with multiple projects underway in Australia and around the world \cite{Passey_CES}. Their feasibility however faces a major barrier from the network charges that are applied to all transactions, including when charging the CES from local solar generation and when discharging the CES to serve local customer demand (the economics of customer DER, in contrast, benefits greatly from displacing customers' network charges, which are a major component of customer total bills \cite{AEMC2021}).


In this paper we establish that the combined deployment of CES and local network charges can resolve these technical and economic challenges by creating economic benefits for all stakeholders and lowering the power demands on the network.
We show that LUOS tariffs can improve the viability of CES and how the addition of CES into a network enables LUOS tariffs to create mutual benefits for all customers and for the DN operator.
Our study focuses on the fundamental interactions between CES and local network charges. We therefore consider the simplified scenario where customers and the CES participate directly in the wholesale energy market. This provides all customers with equal access to the CES (without distinctions of ``participating'' and ``non-participating'' customers) and avoids ambiguities of retail arrangements.
Under these conditions, it is clear that the mutual benefits arise from the CES being able to access greater electricity market arbitrage value when it experiences discounted local network charges. This additional revenue improves the financial viability of the CES while creating additional transactions on the DN that the CES pays network tariffs on. This increase in transaction volume reduces the per unit cost of each transaction on the network (to maintain the revenue of the DN operator). The addition of CES network payments therefore reduces the proportion costs borne by customers.
The improvement in network conditions arises from the network tariff discount being conditional upon power flowing within the local network - either from customers to the CES or from the CES to customers.

The combination of these innovations resolves the barriers that each faces when deployed in isolation, namely that the cost of existing network tariffs hinders the feasibility of CES, and that local network tariffs are generally a zero-sum trade-offs between customer segments and the network operator.
The key criteria for the relationship to be symbiotic is that local network tariffs are set at less than half the price of regular distribution network tariffs. This is because use of the CES incurs twice (or slightly more than twice once accounted for battery losses) the number of local network charges relative to the alternative of importing energy from an upstream grid.
This previously unreported symbiosis brings both CES and local network tariffs closer to mainstream deployment, in a manner that improves the economic and technical outcomes for all stakeholders.
This is especially the case given that the alternative of energy storage on customer's properties reduces the energy throughput of the network. This drives up the per unit energy cost of the network, which is paid for by all customers - primarily by those who don't reduce their energy imports through owning DER.

The remainder of the paper is structured as follows: in Sect.~2 we review the literature of network pricing and CES; in Sect.~3 we introduce our proposed tariff models; in Sect.~4 we outline our simulation method and modelled scenario; in Sect.~5 we present our results, beginning with mathematical criteria, before studying a single illustrative set of tariff values, generalising to examine the effects of a range of tariff values, and discussing the implications of these results; in Sect.~6 we summarise our findings and discuss their policy implications.

\section{Literature Review}
This article builds upon and connects two research fields: the first concerns pricing electricity supply in the presence of large amounts of DER; the second relates to CES.
Traditional models of pricing electricity apply two main types of charges: (1) a network tariff that accounts for the costs associated with transporting electricity from generation to consumption and is typically regulated and applied to all customers serviced by a given network operator, and (2) an electricity price that accounts for the costs of generating electricity and is often set using a market.
Efforts to incorporate energy flows between customers have generally targeted one or the other of these charges.

Proposals to update network tariffs have generally started from the premise that that energy flows between customers ought to be charged a reduced network fee to reflect the lower marginal costs of maintaining a small segment of the network. This was the motivation for the first proposal of a ``Local Use Of Service'' (LUOS) charge by Langham et al. \cite{langham2014} as part of a policy program by the City of Sydney \cite{rutovitz2014level}. This concept has since been applied to both regulated network tariffs \cite{RUTOVITZ2018324} and peer-to-peer local trading schemes, where buyers and sellers could share the network fee \cite{roy2016potential}.

While such proposals are effective at reducing costs for customers, they do so at the direct and proportionate expense of network operators \cite{RUTOVITZ2018324}. This undermines the ability of network operators to recover their costs, making such proposals untenable. 
Such proposals may also have second order effects where the absence/presence of discount tariffs motivates greater/less investments in BTM energy storage. Such storage may on the one hand contribute to reducing energy market peak prices and peak loads on the network, reducing costs for all customers, while on the other hand it reduces the volume of energy flowing on the network, thereby driving up per unit network charges that are borne by customers without their own BTM energy generation and storage.

The other approach, focusing on electricity pricing, is to introduce a separate localised energy market for peer-to-peer (P2P) transactions.
P2P models typically use decentralised markets, such as direct P2P transactions \cite{mengelkamp_role_2017}, closed order books \cite{mengelkamp_trading_2017}, community markets \cite{rodrigues_battery_2020} and coordinated market \cite{tushar_peer--peer_2021}. Some of these market approaches have been extended to incorporate the technical constraints of low voltage networks through optimal power flow calculations \cite{guerrero_local_2019}.
Although there are a number of commercial providers of P2P trading platforms, the deployment of P2P markets has to date been limited as they represent a major departure from conventional regulations and solutions have not been found to address the cost burden of standard network charges being applied to local P2P flows. 

Fundamentally, all of these economic approaches - targeting network prices or electricity prices - are constrained by the same problem: that they operate in a zero-sum environment where the cost of maintaining the distribution network must be recouped from customers. Within this constraint they are limited to redistributing costs and benefits between customer segments.
In this paper we show how the introduction of a CES can break this zero-sum constraint and enable improved outcomes for all stakeholders simultaneously. 

CES is a relatively new type of storage, fitting between small ($~10$kWh) BTM storage systems and large ($~100$MWh) utility systems. The technical and economic performance of CES has been studied under both fixed tariff \cite{barbour_community_2018, dong_impact_2020, arghandeh_economic_2014, talent_optimal_2018, van_der_stelt_techno-economic_2018, terlouw_multi-objective_2019, hafiz_energy_2019, dong_techno-enviro-economic_2020} and dynamic pricing structures \cite{schram_trade-off_2020, walker_analysis_2021}. 
CES assets are in some ways in competition with BTM storage, and many studies have compared their relative merits. When the capacity of CES is optimised, these studies have established that CES can outperform deployments of equivalent sized BTM energy storage systems due to greater energy arbitrage revenues and improved management of peak power imports and exports with figures ranging between 13.8\% and 50\% reduction across studies \cite{barbour_community_2018, dong_impact_2020, talent_optimal_2018, van_der_stelt_techno-economic_2018, terlouw_multi-objective_2019, hafiz_energy_2019, dong_techno-enviro-economic_2020,parra_optimum_2015}. As CES systems are generally envisioned as providing shallow energy storage (of a few hours), studies have determined that Lithium based batteries are the best available technology for CES \cite{terlouw_multi-objective_2019, parra_optimum_2015, parra_optimum_2017}.
Recent CES studies have extended economic optimisations by adding environmental goals and network constraints into CES optimisation algorithm objectives \cite{mediwaththe_network-aware_2021,schram_trade-off_2020,pimm_community_2020,dong_improving_2020}. 
Industry trials, meanwhile, have demonstrated the technical performance of CES as well as some equity concerns. Some trials have made the CES available to only DER owning customers, and even within this cohort found that not all customers benefited from the installation of the CES \cite{Powerbank}.
Like P2P trading models, the viability of CES is significantly encumbered by the burden of network charges being applied to CES transactions, even when these are improving local grid conditions. 
Overcoming this barrier, in a manner that benefits customers as well as network operators, is the central contribution of this paper.

\begin{figure}[]
\centering
\includegraphics[width=0.8\linewidth]{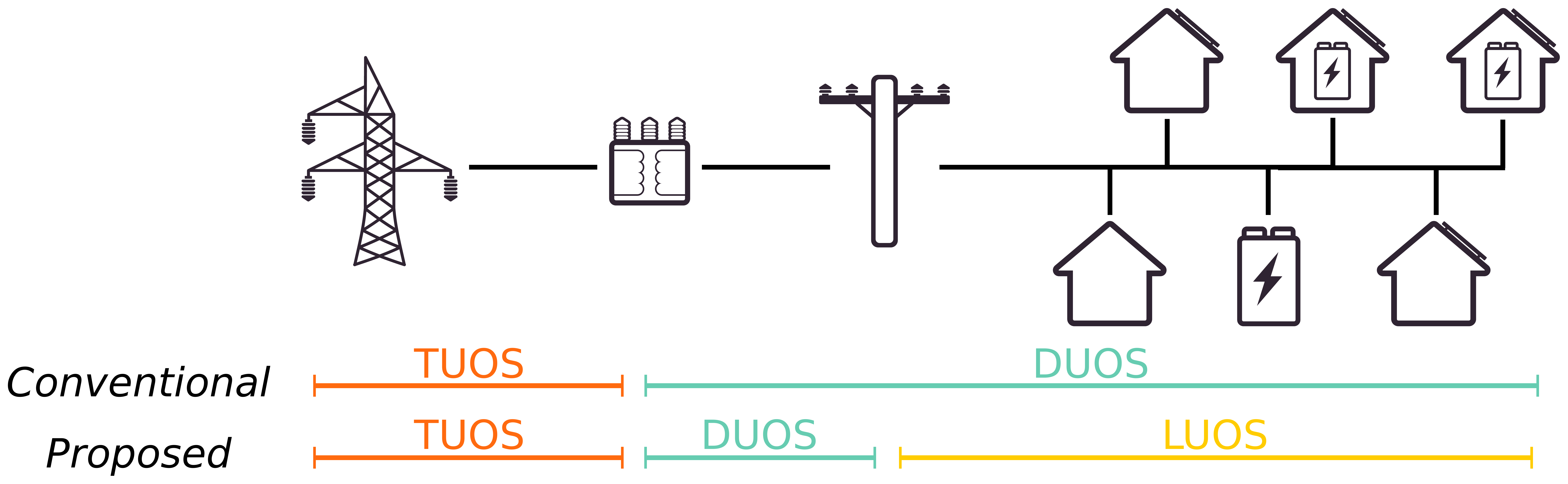}
\caption{Schematic showing the segments of the electricity network. The top line shows the conventional arrangement where TUOS and DUOS charges are distinguished by a distribution substation (transformer icon). The bottom line illustrates and how the proposed LUOS charge replaces the DUOS charge within a localised region of the network (such as below a distribution transformer). HV - High Voltage, MV - Medium Voltage, LV - Low Voltage.
}
\label{tariff_regions}
\end{figure}

\section{Tariff Models}
\label{tariff_sect}
In this paper we adopt a refined distinction between the transmission and distribution systems through a third tier of network pricing for localised regions of the distribution network, as illustrated in Fig.~\ref{tariff_regions}. In this structure, energy flows within a local region are charged a LUOS charge in place of the conventional ``Distribution Use Of Service'' (DUOS) charge, which continues to be levied on flows through the upstream parts of the distribution network (DN).
We assume that there are no reverse power flows into the transmission network and exclude ``Transmission Use Of Service'' (TUOS) charges from our calculations.
The concept of a LUOS charge has investigated by numerous studies \cite{langham2014,rutovitz2014level,roy2016potential}. The distinction of our work is to complement the use of LUOS charges with the deployment of a CES and to tune the interaction between these such that they create a mutual benefit for all stakeholders.
The focus of LUOS on local regions matches the defining physical feature of CES: that CES are located close to customers with energy flows to and from the CES asset utilising only a small fraction of the DN. We consider the boundary between DUOS and LUOS regions to be a Medium Voltage (MV) to Low Voltage (LV) transformer because reductions in the dynamic range of this transformer are likely to offer significant savings in the long run marginal cost of operating the DN. This notwithstanding, our analysis applies equally to an arbitrary boundary.

Unlike previous CES models, which followed the P2P approach of introducing new dynamic energy markets \cite{guerrero_local_2019,mediwaththe2019incentive,mediwaththe2015dynamic,mediwaththe2017competitive,8513887}, our model only affects network tariffs and may be effectively implemented using static tariffs - that are defined far ahead of time and are applicable to all connection points within a region, such as those typically set by regulators - rather than dynamic tariffs that change based on real time market dynamics.
We consider the CES to be subject to the same network tariffs as customers (as both connect to the same voltage level of the network). We also consider the CES and customers to have direct access to the electricity market (rather than this relationship being mediated by retailers) as this captures the underlying cost of their electricity supply (and avoids ambiguity in retail pricing).
Customers' direct access to the singular market price furthermore avoids any distinction between customers who are ``participating'' in trading with the CES and/or local customers and those that are ``non-participating'', as all customers experience the same prices at each moment. The customers also all experience the same network prices, which are applied throughout the local network area at each moment in time.

In addition to introducing local tariffs, we explore the impact of applying DN tariffs on exported energy, as well as imported energy, as is currently being (contentiously) considered by the Australian Energy Market Commission \cite{AEMC2021}. Distinguishing between imports and exports, as well as between charges for network services and electricity generation, defines a total of eight distinct charges.
We represent each of these charges with the symbol $\lambda$ and use the following nomenclature to capture their details: superscripts $e$ demarcate charges that covers the cost of electricity generation as set by an electricity market, while superscripts $n$ demarcate network charges that cover the cost of transporting electricity in the network.
Subscripts denote the direction ($+$ for imports, $-$ for exports) and the region of flows ($l$ for flows within the local area and $u$ for flows to/from the upstream grid).
The eight charges are summarised in Table~\ref{table:1}.

\begin{table}[]
\centering
\begin{tabular}{|c|c|c|c|c|} 
 \hline
 & \multicolumn{2}{c}{upstream} & \multicolumn{2}{|c|}{local}\\
 \hline
           & import & export & import & export \\ \hline
 energy    & $\lambda^e_{u+}$ & $\lambda^e_{u-}$ & $\lambda^e_{l+}$ & $\lambda^e_{l-}$ \\ \hline
 network & \cellcolor{cyan!25} $\lambda^n_{u+}$ &  \cellcolor{cyan!25} $\lambda^n_{u-}$ &  \cellcolor{yellow!25} $\lambda^n_{l+}$ &  \cellcolor{yellow!25} $\lambda^n_{l-}$ \\ 
 \hline
\end{tabular}
\caption{The eight charges that apply within the proposed tariff models. Charges cover the cost of electricity generation as set by an electricity market (e) and the cost of transporting electricity in the network (n). Our study excludes the transmission network such that upstream network charges (highlighted in blue) are DUOS tariffs ($\lambda^n_{u+}, \lambda^n_{u-}$) and local network charges (highlighted in yellow) are LUOS tariffs ($\lambda^n_{l+}, \lambda^n_{l-}$).}
\label{table:1}
\end{table}

The focus of this paper is to understand the impact of introducing LUOS tariffs, and on making DUOS and LUOS charges applicable on exports as well as imports. To explore these modifications we study three network tariff models, which we refer to as follows:
\begin{enumerate}
    \item {\bf DUOS} - this is the business as usual case where all imports of electricity are charged DUOS rates (there is no distinction for LUOS), while exports do not incur any network charges: $\lambda_{l+}^n = \lambda_{u+}^n$; $\lambda_{l-}^n = \lambda_{u-}^n = 0$.
    \item {\bf 1-way LUOS} - this model distinguishes flows within the local network from those between the local network and the upstream network. Connection points importing locally produced electricity are charged at LUOS rates, while electricity imported from the upstream network is charged the DUOS rate. Electricity exports continue to be excluded from charges: $\lambda_{l-}^n = \lambda_{u-}^n = 0$.
    \item {\bf 2-way LUOS} - in this model DUOS and LUOS charges are both applied to electricity exports as well as imports.
\end{enumerate}

Throughout our study we consider network tariffs to be fixed, flat rates that apply uniformly throughout each day (as opposed to `Time of Use' (ToU) tariffs that vary across the day). These are charged on a per kWh basis, with fixed daily connection charges excluded in the interest of clarity. Electricity prices, meanwhile, are considered to be set by an energy market that is truly two-sided, with electricity prices applying equally to imports and exports from each connection point: $\lambda^e = \lambda^e_{u+} = \lambda^e_{u-} = \lambda^e_{l+} = \lambda^e_{l-}$. We assume the CES and customers are pure price takers in this market, with their demand and generation not affecting prices.

\section{Simulation Method}

To assess the outcomes of the different tariff models we simulate the behaviour of a CES system connected within a local network of 100 households. 
Of the 100 households, a randomly selected 50\% are taken to have solar photovoltaic systems.
This high solar penetration level is motivated by the Australian context where dozens of postcodes already have penetration levels of 50\% \cite{CC2017}, exceeding the Zero Carbon scenarios considered in previous work \cite{parra_optimum_2015} and even the 40\% considered in more recent studies \cite{barbour_community_2018}. 
Higher and lower penetration levels were explored and found to not fundamentally alter the behaviour of the CES, although, as noted in previous CES studies \cite{barbour_community_2018}, larger amounts of locally generated solar energy enhance the positive impacts of CES.

We considered the customers and CES to be situated within a bounded local subsection of a DN. This local network is connected to a larger upstream power system through an unconstrained connection, which allows energy to be imported and exported from the local network as needed. We do not consider the topological details of the distribution network, focusing on the net energy flows between the local network and the upstream grid, rather than on power flows within the local network - these will be the subject of further work.

To enhance the practical impact of our study, we utilise real world data for prices and power. We locate our simulations in the state of South Australia for three reasons: it has Australia's highest solar penetration at over 33\% \cite{energymag}, it has a volatile electricity market that is well suited to CES arbitrage, and the local DN operator is actively pursuing regulatory approval to introduce 2-way network tariffs that apply to customers exporting power into the DN \cite{AEMC2021}. We use electricity market price data from Australian Energy Market Operator \cite{nemweb} and DUOS charges are taken from South Australian Power Networks \cite{SAPN}.
High quality data for Australian household electricity demand and solar generation (5-minute resolution) was taken from the Nextgen trial \cite{Shaw2019}. While this trial took place in another region (the Australian Capital Territory) the data is likely to be representative of solar owning households in South Australia, especially as the population distributions and electricity systems of each region are dominated by cities that lie at very similar latitudes (within $0.36^{\rm o}$). A more systemic bias in the Nextgen data set, as well as many fine-grained household data sets \cite{Kapoor2020}, is that the households are likely all financially well off (as participation in the trial is predicated on installation of solar and a home battery), which has significant impacts on how they use electricity through factors such as the quality of their houses and appliances and their lifestyles \cite{Ransan-Cooper2021}. 

\begin{figure}[]
\centering
\includegraphics[width=0.5\linewidth]{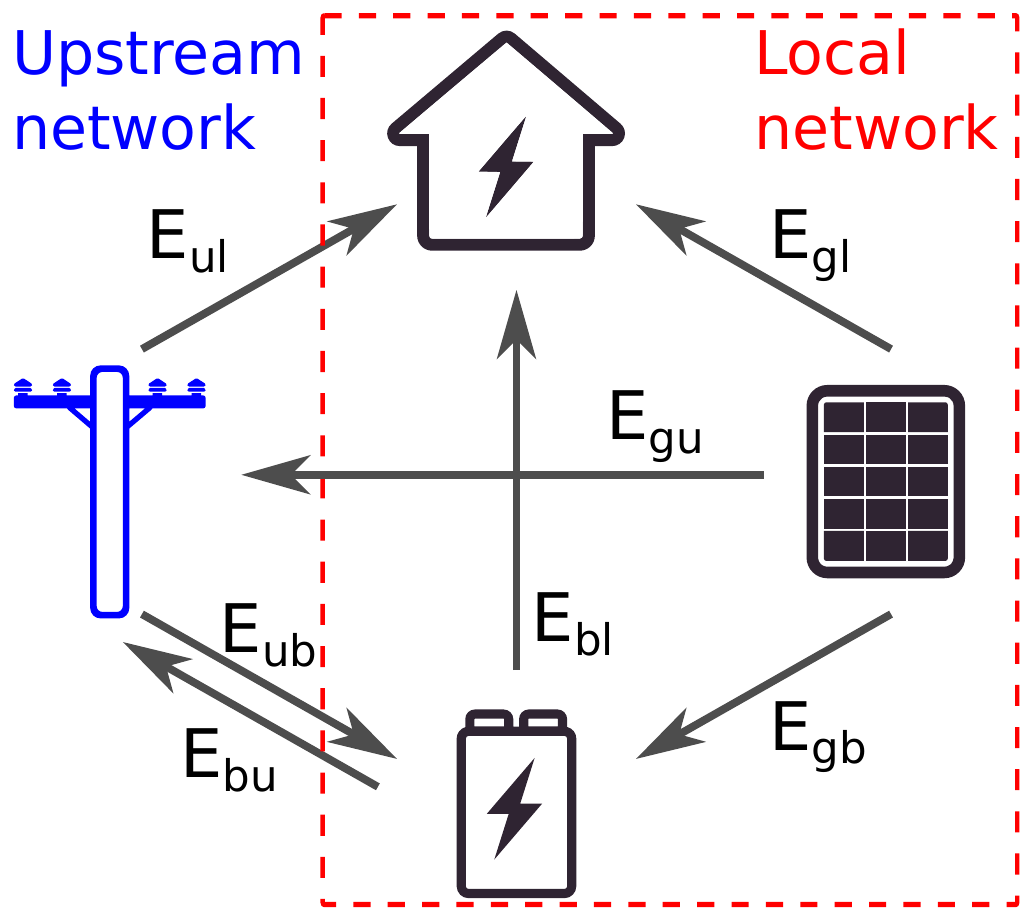}
\caption{
Schematic showing the seven possible flows of energy between the constituents of a local network segment (assets contained within red dashed box) and an upstream network (distribution pole icon referred to as ``u'' for upstream). 
At each instance of the calculation we divide households into two groups: those with excess solar generation that is flowing into the DN, who we represent with a solar panel icon and label ``g'' for generation, and those with greater load than BTM solar generation, who we represent with a house icon and label ``l'' for load. We sum the energy generation/load values together for each group to get the total energy flowing through the DN. Energy flowing between generating households to consuming households is labelled E$_{\rm gl}$, energy flowing from these households to the upstream grid is E$_{\rm gu}$ and energy flowing to the CES is E$_{\rm gb}$.
The CES is indicated by a battery icon and ``b'' labels, although it could be any technology. The CES can charge from the upstream grid (E$_{\rm ub}$) or households' excess solar energy generation (E$_{\rm gb}$), and can discharge to the net load households (E$_{\rm bl}$) or the upstream network E$_{\rm bu}$.
Energy flowing from the upstream network to households with net loads is labelled E$_{\rm ul}$.
}
\label{framework}
\end{figure}

The total generation capacity of the 50 household solar systems is 268kW and the CES was assigned an energy capacity of 380kWh, which is equal to the total capacity of the BTM batteries installed in the 50 Nextgen households that we include as solar households in this simulation. The CES is chosen to have two hours of storage, so the power rating is 190kW.
Its actions are determined by an optimisation algorithm whose objective is to minimise the aggregate electricity costs of the 100 households and the CES. 
Pursuing this objective requires knowledge of the net load of the community, which in the absence of the CES would flow from/into the upstream network. Such data could be provided to the CES through a single meter installed at the connection point between the local and upstream networks. 

Modelling the operation of the CES and the customers involves a consideration of a total of seven possible energy flows, as represented in Fig.~\ref{framework}. In this model, each household is, at each time interval, classified as either a net generator of power, if their solar generation exceeds their BTM load (depicted as a solar panel in Fig.~\ref{framework}) or a net load, if their load is greater than their solar generation (depicted as a house in Fig.~\ref{framework}).
From these groupings, the net energy flows between net generation and net load households (labelled E$_{\rm gl}$) can be calculated, together with any excess net generation that may flow either to the upstream grid (labelled E$_{\rm gu}$) or to the CES (labelled E$_{\rm gb}$). 
Only energy flows that both originate and terminate within the local network, indicated by the red dashed box in Fig.~\ref{framework}, are eligible for a LUOS network tariff ($\lambda_{l}^n$). These are E$_{\rm gl}$, E$_{\rm gb}$ and E$_{\rm bl}$.
All flows between the upstream grid (depicted by the power pole in Fig.~\ref{framework}) and the local network are charged a DUOS rate ($\lambda_{u}^n$).

The objective of the CES algorithm is set to minimise the cost for households and the CES, which, in the two-sided market with equal prices for buying and selling electricity ($\lambda_{l+}^e = \lambda_{l-}^e$) is:
\begin{equation}\label{customer_total}
\begin{split}
C_{\rm Cust + CES} = (\lambda_{u+}^e + \lambda_{u+}^n)(E_{ul}+E_{ub})
                    -(\lambda_{u-}^e - \lambda_{u-}^n)(E_{gu}+E_{bu})
                    +(\lambda_{l+}^n + \lambda_{l-}^n)(E_{gl}+E_{gb}+E_{bl}).
\end{split}
\end{equation}

This optimisation constitutes a linear problem that we solve using our open source python module, called c3x \cite{c3xgithub}. 
As the focus is on the impact of different tariffs on CES operation we provided the algorithm with perfect foresight of market prices and the net load of each household across the simulation period.
For further details of the optimisation implementation see Appendix \ref{appendix}.

To assess the impacts of the different tariff models on other stakeholders we also define the collective cost to customers in Eq.~\ref{customer_cost}, the cost (negative costs are revenues) of the DN operator in Eq.~\ref{network_cost} and the cost of the CES operator in Eq.~\ref{battery_cost}.

\begin{equation}\label{customer_cost}
\begin{split}
C_{\rm Customers} = (\lambda_{u+}^e + \lambda_{u+}^n)E_{ul} 
                   +(\lambda_{l+}^n + \lambda_{l-}^n)(E_{gl})
                   +(\lambda_{l+}^e + \lambda_{l+}^n)(E_{bl})
                   -(\lambda_{l-}^e - \lambda_{l-}^n)E_{gb} 
                   -(\lambda_{u-}^e - \lambda_{u-}^n)E_{gu}.
\end{split}
\end{equation}

\begin{equation}
\begin{split}
C_{\rm DN} = -\lambda_{u+}^n(E_{ul}+E_{ub})
             -\lambda_{u-}^n(E_{gu}+E_{bu})
             -(\lambda_{l+}^n + \lambda_{l-}^n)(E_{gl}+E_{gb}+E_{bl}).
\label{network_cost}
\end{split}
\end{equation}

\begin{equation}\label{battery_cost}
\begin{split}
C_{\rm CES} = (\lambda_{u+}^e + \lambda_{u+}^n )E_{ub} +
                    (\lambda_{l+}^e + \lambda_{l+}^n)E_{gb} -
                    (\lambda_{u-}^e - \lambda_{u-}^n)E_{bu} -
                    (\lambda_{l-}^e - \lambda_{l-}^n)E_{bl} .
\end{split}
\end{equation}

We also assess the impacts in terms of the self-sufficiency of the local network (Eq.~\ref{self-sufficiency})
as well as the self-consumption of the network's solar generation within the network (Eq.~\ref{self-consumption}). 

\begin{equation}
{\rm self-sufficiency} = 1 - ({\rm E}_{\rm ul}+{\rm E}_{\rm ub}) /
({\rm E}_{\rm ul}+{\rm E}_{\rm ub} + {\rm E}_{\rm gl}+{\rm E}_{\rm gb}),
\label{self-sufficiency}
\end{equation}

\begin{equation}
{\rm self-consumption} = ({\rm E}_{\rm gl}+{\rm E}_{\rm gb}-{\rm E}_{\rm gbu}) /
({\rm E}_{\rm gl}+{\rm E}_{\rm gb}+{\rm E}_{\rm gu}),
\label{self-consumption}
\end{equation}
where ${\rm E}_{\rm gbu}$ is any flow from the generation to the upstream grid via the battery:
\begin{equation}
{\rm E}_{\rm gbu} = 
\begin{cases}
    {\rm E}_{\rm bu}-{\rm E}_{\rm ub},& \text{if } {\rm E}_{\rm bu}-{\rm E}_{\rm ub} \geq 0\\
    0,              & \text{otherwise}.
\end{cases}
\end{equation}

\section{Results and Discussion}

Our aim is to understand how local network tariffs shape the behaviour of CES systems and how this interaction can be arranged to be symbiotic, so as to provide positive financial and technical impacts for all customers and the DN operator.
We achieve this aim in three stages. 
Firstly, we derive the criteria that define the conditions under which it is viable for the CES to perform a charge-discharge cycle. These reveal that there are two mechanisms by which the CES can achieve its goal, and provide guidance on the relative value of DUOS and LUOS that facilitate this goal.
Secondly, we study the behaviour of the CES under a specific set of tariff values, examining the charge/discharge actions of the CES in detail across a couple of days. 
Thirdly, we survey the behaviour of the CES across a wide range of DUOS and LUOS tariff values. 
These surveys demonstrate that there exist a range of DUOS and LUOS tariff values that lead to mutually beneficial outcomes for customers and DN operators, relative to current (DUOS only) network tariffs.

\subsection{Defining Viable CES Charge-discharge Cycles}
\label{results0}
There are two avenues by which the CES can reduce the collective costs of the CES and customers.
\begin{enumerate}
    \item The CES can earn a profit from arbitraging the electricity market, buying when it is cheap and selling when it is expensive.
    \item The CES can reduce customers' bills. Under the conditions that there is a single price for electricity, which applies equally to imports and exports and is unaffected by CES actions, the CES is unable to alter the energy component of customers' bills. (These are always simply the product of each household's net load profile with the electricity market price.)
    The CES can however reduce customers' network charges (if LUOS < DUOS) by increasing the amount of electricity that circulates within the local network rather than being transacted with the upstream network.
\end{enumerate}
Both of these actions (and indeed any actions by the CES) increase the number of energy transactions in the DN and thereby introduce additional network charges. When the CES buys and sells energy from the upstream network these energy flows are clearly additional. When the CES charges from local solar generation and later discharges to provide energy to local customers there are four potential network charges (two of these are zero in the 1-way LUOS tariff model): the CES is charged $\lambda_{l+}^n$ for charging and $\lambda_{l-}^n$ for discharging, while customers are charged $\lambda_{l-}^n$ for exporting and $\lambda_{l+}^n$ for importing. In the comparative case where customers export and later import from upstream, there are half as many potential network charges: customers are charged $\lambda_{u+}^n$ for exporting (which is zero in the 1-way LUOS tariff model) and $\lambda_{u-}^n$ for importing.

These additional network charges, which are inescapable for devices connected directly into the network (rather than BTM), create an interdependence between the two avenues by which the CES can achieve its objective.
While the precise criteria defining these interdependencies depend on the network tariff model (detailed below), in all cases there exists a threshold when paying twice the number of LUOS charges is cheaper than the DUOS charges of exchanging energy with the upstream network.

Understanding the two avenues by which the CES can derive value also explains how the CES breaks the typical zero-sum trade-offs between stakeholders that normally constrain the viability of tariff modifications. This is made possible by the CES deriving new value from its arbitrage of the energy market. Some of this value can flow to the DN operator, who records a greater number of transactions and therefore accepts reduced LUOS rates. This is consistent with current regulated network pricing structures, where higher DN energy throughput reduces the per unit cost of network services. This transfer from the CES to the DN alleviates customers from the full burden of covering the DN operator's revenue.
This mechanism suggests that stakeholders could benefit further if the CES were to access additional revenue streams, such as participating in frequency support markets.

\subsubsection{DUOS Model}
Under the DUOS model the CES is unable to reduce customers' network charges but is still able to derive revenue from energy market arbitrage. Under these conditions, a profitable charge-discharge cycles requires:
\begin{equation}
\lambda_{sell}^{e} - \lambda_{buy}^{e} > \lambda_{u+}^{n} + \$_\text{throughput},
\label{duos-arbitrage}
\end{equation}
where 
$\lambda_{sell}^{e}, \lambda_{buy}^{e}$ are the electricity prices at the time the CES discharges and charges respectively and 
$\$_\text{throughput}$ is the per kWh cost of cycling the CES, such as the degradation of a battery.

\subsubsection{1-way LUOS Model}
Under the 1-way LUOS model, both avenues for deriving benefits are open to the CES and a viable charge-discharge cycle is defined by:
\begin{equation}
\lambda_{sell}^{e} - \lambda_{buy}^{e} > 2\lambda_{l+}^{n} - \lambda_{u+}^{n} + \$_\text{throughput}.
\label{luos-arbitrage}
\end{equation}

\subsubsection{2-way LUOS Model}
Under the 2-way LUOS model, the comparison remains the same as in the 1-way LUOS model, but the number of network charges levied on transactions doubles as they are charged on imports and exports. In this case profitability requires:
\begin{equation}
\lambda_{sell}^{e} - \lambda_{buy}^{e} > (2\lambda_{l+}^{n} + 2\lambda_{l-}^{n}) - (\lambda_{u+}^{n} + \lambda_{u-}^{n}) + \$_\text{throughput}.
\label{2-way-arbitrage}
\end{equation}

Equations~\ref{luos-arbitrage},\ref{2-way-arbitrage} show how the introduction of discount LUOS tariffs ($\lambda_{l}^{n} < \lambda_{u}^{n}$) decreases the threshold the CES needs to clear to derive profitable energy market arbitrage. Furthermore, once $2\lambda_{l}^{n} < \lambda_{u}^{n} - \$_\text{throughput}$ it becomes possible for the CES to produce a net benefit even when there is no energy arbitrage opportunity $\lambda_{sell}^{e} \leq \lambda_{buy}^{e}$.

\subsection{Illustrative Tariff Values}
\label{results1}

To see how the criteria for viable CES cycles play out under each tariff model we study the operation of the CES across the month of January.
We base our case study in the state of South Australia, adopting real historic values for the electricity spot market \cite{nemweb} and (DUOS) network charges from the South Australian DN operator \cite{SAPN}. The throughput cost is set to $\$_\text{throughput} = 3.2$c/kWh, based on the warrantied cycles of batteries \cite{throughput}.
For the 1-way LUOS and 2-way LUOS models we adopt network tariff values that maintain a constant revenue for the DN operator (within 2\% of revenue under current tariffs without the CES, as well as 2\% of current DUOS only tariffs with the CES). This is done because significantly reduced revenue will be untenable for the DN to cover the cost of maintaining the network, while major increases will be unacceptable to customers. 
Our chosen tariff values are:
\begin{enumerate}
    \item {\bf DUOS} - all imports of electricity are charged the DUOS tariff of 13.2c/kWh, while exports do not incur any network fees: $\lambda_{u+}^n = \lambda_{l+}^n = 13.2$c/kWh, $\lambda_{l-}^n = \lambda_{u-}^n = 0$c/kWh.
    \item {\bf 1-way LUOS} - the LUOS rate is discounted to $\lambda_{l+}^n = 4$c/kWh, which requires the DUOS rate to be increased to $\lambda_{u+}^n =15$c/kWh to maintain DN operator revenue. Exports are exempt from network charges ($\lambda_{l-}^n = \lambda_{u-}^n = 0$c/kWh). This presents the CES with the opportunity to save 7c/kWh round-trip on network charges by keeping energy within the local network: $2\lambda_{l+}^{n} - \lambda_{u+}^{n} = 7$c/kWh in Eq.~\ref{luos-arbitrage}.
    \item {\bf 2-way LUOS} - both DUOS and LUOS network charges are applied equally on imports and exports. We set the DUOS tariff to $\lambda_{u+}^n = \lambda_{u-}^n = 13.2$c/kWh (as in DUOS model) and the LUOS tariff to $\lambda_{l-}^n = \lambda_{l+}^n = 4.85$c/kWh. In addition to maintaining the revenue of the DN operator, these tariffs also present the CES with the same potential saving in round-trip network charges as in the 1-way LUOS model: $(2\lambda_{l+}^{n} + 2\lambda_{l-}^{n}) - (\lambda_{u+}^{n} + \lambda_{u-}^{n}) = 7$c/kWh in Eq.~\ref{2-way-arbitrage}.
\end{enumerate}

\begin{figure}[]
\centering
\includegraphics[width=0.7\linewidth]{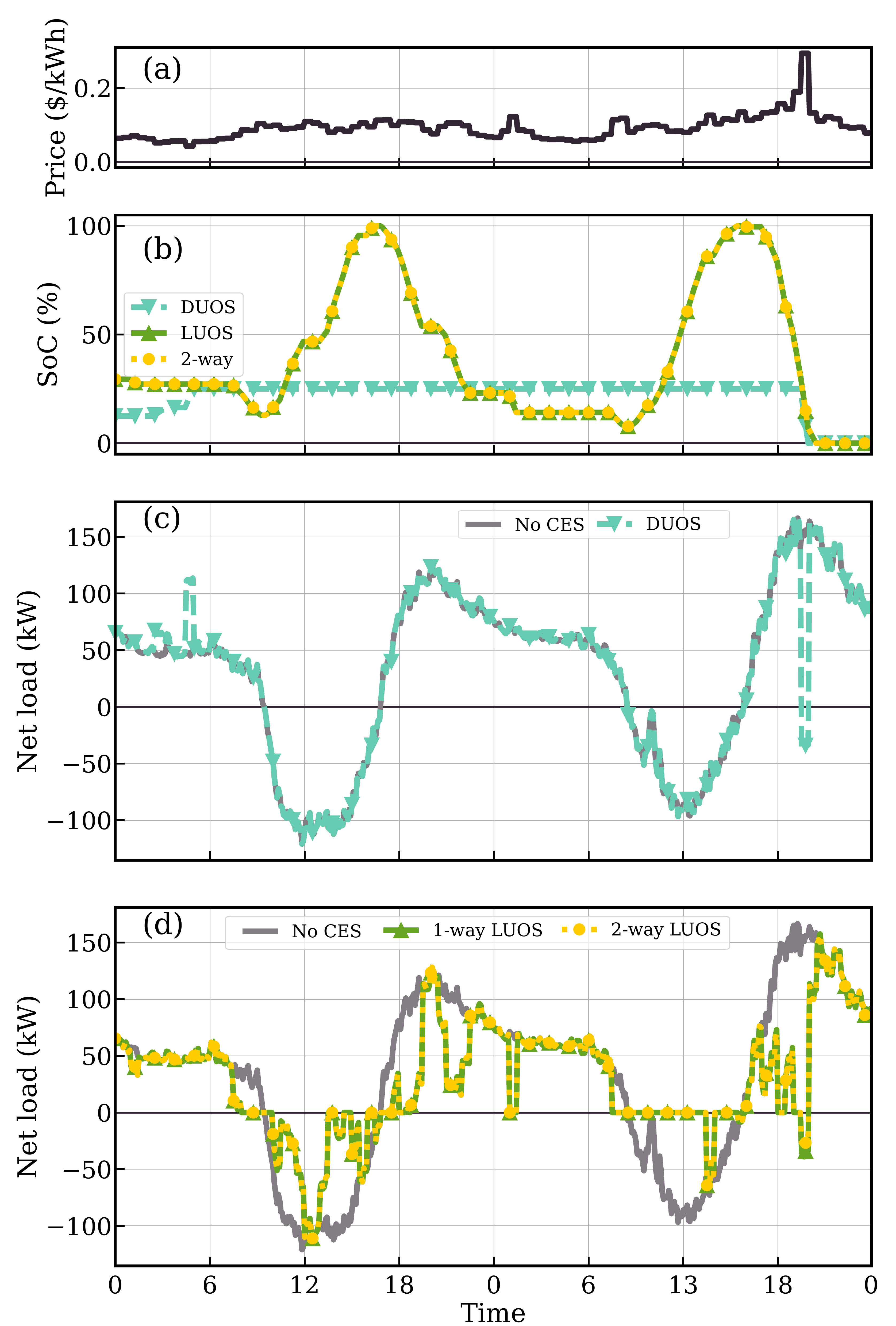}
\caption{(a) The electricity market prices in South Australia across the 3rd and 4th of January 2018.
(b) The SoC of the CES when operating under the three different tariff models. The results are a snapshot of a month long optimisation, which means the CES State of Charge (SoC) at the beginning of the time window varies under different models. (c) The net load of the local network without the CES and under the DUOS model. (d) The net load of the local network under the LUOS models, with the case without the CES shown for comparison.
}
\label{days}
\end{figure}

Figure~\ref{days} shows how the LUOS tariff models incentivize significantly different CES charge/discharge actions. The figure presents results for the 3rd and 4th of January 2018, showing the wholesale electricity price ($\lambda_{u+}^{e} = \lambda_{u-}^{e} = \lambda_{l+}^{e} = \lambda_{l-}^{e}$) in (a) together with the State of Charge (SoC) of the CES in (b), and the net load of the local network on the upstream connection point under the different tariff models in (c) and (d), together with the reference case where there is no CES.

Under the DUOS model the criteria for profitable arbitrage is met rarely. The CES only charges during a short period of low prices on the first day and discharges at the peak price period on the second day (see Fig.~\ref{days}(b)). In both of these periods the actions of the CES increase the strain that the local networks places on the upstream grid. As shown in Fig.~\ref{days}(c)), the CES firstly increases the net load and then creates additional energy exports. This reflects that the CES has no incentive to consider the availability of local generation or demand.

The introduction of local network tariffs in the 1-way LUOS and 2-way LUOS models links the operation of the CES with the net load of the local network. Figure~\ref{days} demonstrates how our choice of setting an equal difference in round-trip network charges in the models leads the CES to perform precisely the same charge/discharge actions. Figure~\ref{days}(c) shows how the discount LUOS incentivises the CES to charge from excess local solar generation (when net load is negative) and discharge to cover local demand because both these actions expose the CES and customers to LUOS charges rather DUOS network charges.
The price peak on the second day is so great that it continues to be profitable for the CES to discharge energy into the upstream network, despite the larger DUOS network charges for doing so.
This results in a much greater utilisation of the CES, which in Fig.~\ref{days}(b) is shown to undergo almost a full charge-discharge cycle each day.
Figure~\ref{days}(b) also shows how the savings in round-trip network charges (7c/kWh) makes it viable for the CES to absorb much, but not all, of the available solar exports. This is because the CES can accept losses in the energy market arbitrage of up to, but no more than 3.8c/kWh ($\lambda_{sell}^{e} - \lambda_{buy}^{e} > -7\text{c/kWh} + \$_\text{throughput}$).

The cumulative effect of the LUOS tariffs driving greater CES utilisation is that they increase the self-sufficiency of the local network (Eq.~\ref{self-sufficiency})
as well as the self-consumption of the network's solar generation within the network (Eq.~\ref{self-consumption}). 
Across the month of January, the self-sufficiency increases from 42\% without the CES to 48\% under the LUOS tariffs. This is in contrast with the CES operating under DUOS only tariffs, which decreases the self-sufficiency by 1\% through arbitrage transactions with the upstream network.
Similarly, the self-consumption is 81\% without the CES, 79\% under DUOS only tariffs, and 93\% under LUOS tariffs.

\begin{figure}[]
\centering
\includegraphics[width=0.65\linewidth]{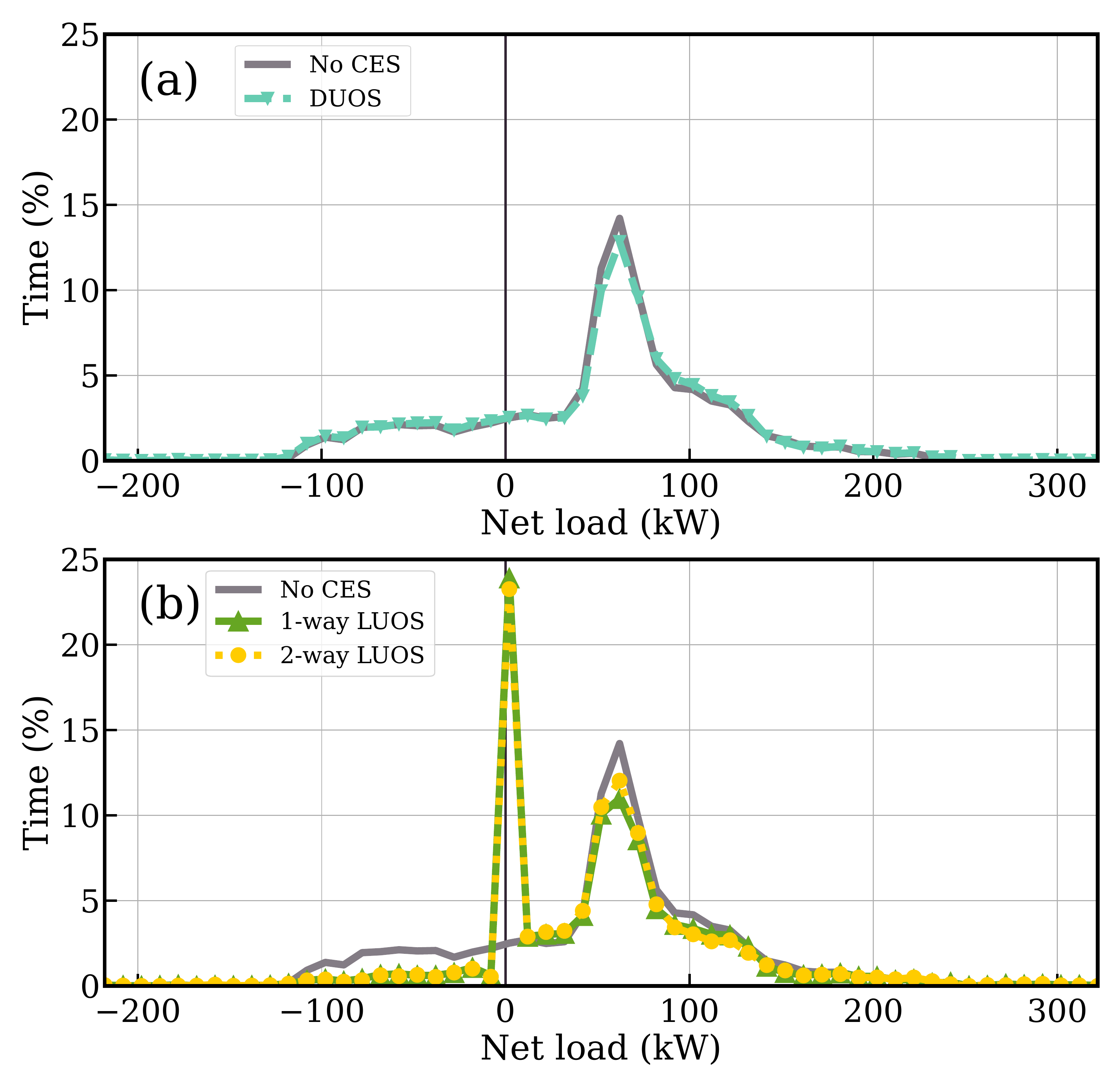}
\caption{
Distributions of time intervals during which the net power flow between the local network and upstream network is positive (importing energy from the upstream network) and negative (exporting power to the upstream network). Distributions are calculated across the month of January for each tariff model and the case where there is no CES.
(a) The distributions without the CES and under the DUOS model, (b) the distributions under the LUOS models, with the case without the CES shown for comparison.}
\label{distribution}
\end{figure}

To examine the impacts of the CES actions on the technical performance of the network we calculate the distribution of the average power flows between the local network and the upstream network (ie. the net load) during each 5-minute data interval in the month. Figure~\ref{distribution} shows the distributions without a CES, under the DUOS model, and the 1-way and 2-way LUOS models. Positive values indicate flows into the local network and negative values indicate flows out of the local network. 
The distribution is centred around a net load of roughly 65kW, represents the average base load of households consuming 650W. 
The distribution under the DUOS model is negligibly different from the case without a CES, reflecting the infrequent actions of the CES in this case. The LUOS models meanwhile significantly decrease the occurrence of power exports and increase the occurrence of self-sufficiency of the local network, particularly increasing the time at which the net load is zero from 2.5\% in the absence of the CES or a CES operating under the DUOS model, to 23\% under the LUOS models.

The results of this section demonstrate how LUOS is critical for increasing utilisation of CES, driving increased self-consumption and self-sufficiency, and reducing the stress on the interconnection between the local and upstream networks.


\FloatBarrier
\subsection{Generalisation Across Tariff Values}
\label{results2}

Building on the findings for a specific set of network tariff values in the preceding section, we now examine how stakeholder outcomes vary as a function of DUOS and LUOS prices. We do this by optimising the CES behaviour under DUOS prices ranging from 12c/kWh to 20c/kWh and LUOS prices ranging from 0c/kWh to 14c/kWh. For each tariff, we simulate a six month period - January through June - so as to cover seasonal variation in solar generation and electricity demand.

\subsubsection{Identifying Mutually Beneficial Tariff Settings}
While in the absence of the CES there is an inescapable trade off between customer costs and network revenue, we show that the introduction of a CES, whose behaviour depends on the network tariffs, creates a dynamic through which reduced LUOS tariffs simultaneously benefit customers and the network operator. 
Mutual benefits occur when the LUOS tariff is decreased and the DUOS tariff is slightly increased, relative to the reference case. For solar and non-solar customers, the reduction of network tariffs applied to imports/exports from/to their neighbours and the CES outweighs the increased network tariffs on upstream imports/exports. For the CES, the reduced network tariffs on energy procured from local solar exports and delivered to local demand enables a greater number of profitable arbitrage trades on the electricity market.
For the network, the increased volume of flows on the local network (into and out of the CES) outweigh the reduction in per unit network charges on these local flows. These counteracting forces are summarised in Table~\ref{table:2}.

\begin{table}[]
\centering
\begin{tabular}{|l|c|c|c|} 
 
 \multicolumn{4}{l}{\bf{(a) 1-way LUOS}} \\ \hline
 \bf{Stakeholders} & \bf{Loses} & \bf{Constants} & \bf{Gains} \\ \hline
 \bf{Solar}    & Increased DUOS & Electricity profile & Discounted LUOS \\ 
     & & Electricity price &  \\ \hline
 \bf{Non-solar} & Increased DUOS & Electricity profile & Discounted LUOS \\
  & & Electricity price & \\ \hline
 \bf{CES}    & Increased DUOS & Availability of local generation & Discounted LUOS \\
  &  & Availability of local load & Increased utilisation \\
  &  & & Increased arbitrage revenue \\ \hline
 \bf{Network} & Discounted LUOS & Total revenue & Increased DUOS \\ 
 & & & Increased volume of flows \\ 
 & & & Reduced peak loads \& exports \\ \hline
\hline
 
 \multicolumn{4}{l}{\bf{(b) Additions from 2-way LUOS}} \\ \hline
 \bf{Stakeholders} & \bf{Loses} & \bf{Constants} & \bf{Gains} \\ \hline
 \bf{Solar}    & DUOS \& LUOS on exports & &  \\ \hline
 \bf{Non-solar} & & & Increased discount on LUOS imports \\ \hline
 \bf{CES}    &  &  &  \\ \hline
 \bf{Network} & & &  \\ \hline 
\end{tabular}
\caption{A summary of the ways in which the introduction of LUOS tariffs influence each stakeholder, creating some loses and gains, and leaving some things unchanged. Table (b) shows the additional changes introduced by moving from the 1-way LUOS model to the 2-way LUOS model.}
\label{table:2}
\end{table}

\begin{figure}[]
\centering
\includegraphics[width=0.98\linewidth]{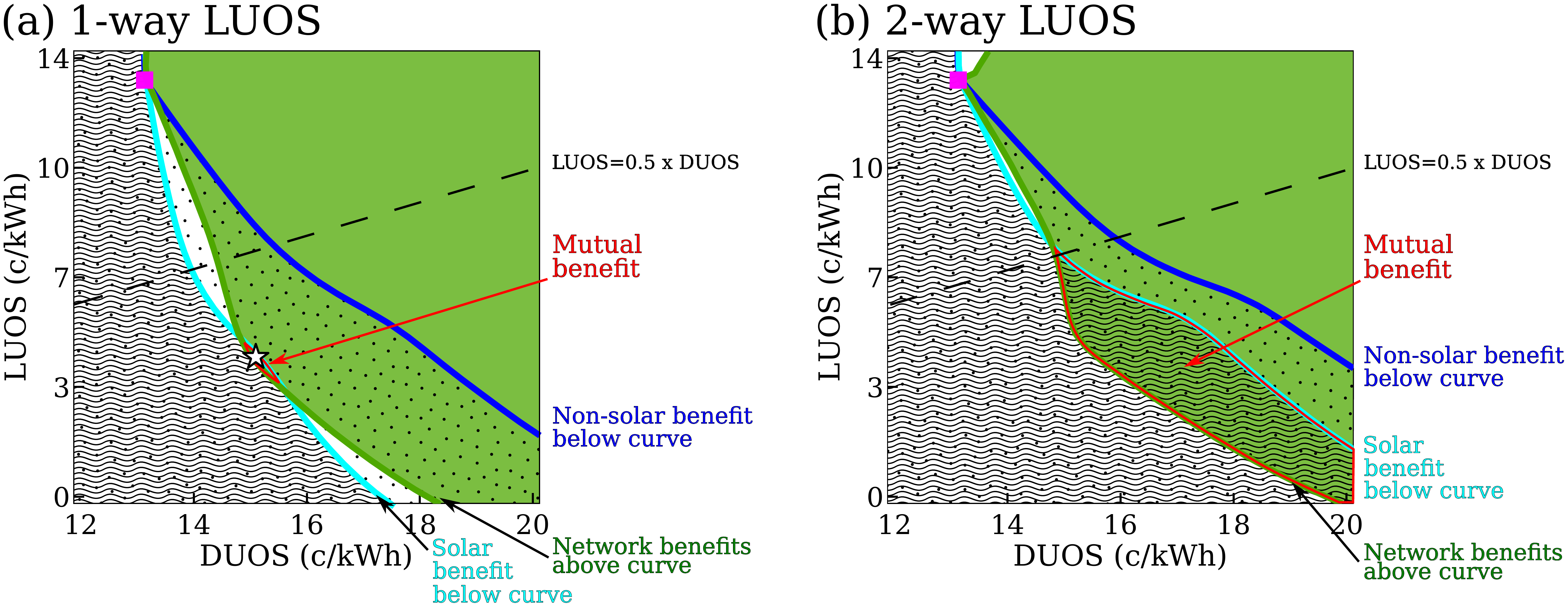}
\caption{
Identification of the range of DUOS and LUOS tariffs under which each stakeholder group is better off than under a reference case (indicated by the purple square). (a) Under the 1-way LUOS model and (b) under the 2-way tariff model. The green curves indicate where the DN operator obtains equal revenue to the reference case. At all tariff combinations above this curve - shaded green - the DN operator receives greater revenue than in the reference case. 
The dark blue curves indicate where non-solar customers pay equal network fees to the reference case. At all tariff combinations below this curve - marked with polka dots - these customers' costs are reduced, relative to the reference case. 
The light blue curves indicate where solar customers pay equal network fees to the reference case. At all tariff combinations below this curve - marked with thin wavy lines - these customers' costs are reduced, relative to the reference case. 
Crucially, we note that under both tariff models there exist a range of DUOS, LUOS tariff values where the DN operator, non-solar customers, and solar customers are all better off than in the reference case. This area is highlighted by the red curves.
The dashed line marks where LUOS~$= 0.5 \times$~DUOS.
The white star in (a) indicates the tariff values studied in Sect.~\ref{results1} (DUOS = 15c/kWh and LUOS = 4c/kWh). The 2-way example from Sect.~\ref{results1} is not directly transferable to (b) because the ratio of import to export tariffs is different.
}
\label{combine}
\end{figure}

The balancing of these effects is demonstrated by Figs.~\ref{combine}(a),(b), where we indicate with different shadings the pricing regions in which solar owning customers, non-solar customers, and the DN operator are each better off than under conventional conditions. Crucially, these figures show that there are a range of DUOS and LUOS prices where all three stakeholder groups are better off (where all three shadings overlap).

Figure.~\ref{combine}(a) shows the results for the 1-way LUOS model, where there are no network charges for exported power. The costs/revenues for stakeholders are compared to the same business as usual reference case of existing South Australian network tariffs used in the preceding sections ($\lambda_{u+}^n = \lambda_{l+}^n = 13.2$c/kWh, $\lambda_{l-}^n = \lambda_{u-}^n = 0$c/kWh).
Figure~\ref{combine}(b) meanwhile, presents the results for the 2-way LUOS model, where we set the reference case to have the same network charges for import ($\lambda_{u+}^n = \lambda_{l+}^n = 13.2$c/kWh) with an additional charge of 2c/kWh on exports ($\lambda_{l-}^n = \lambda_{u-}^n = 2$c/kWh).
This matches the network charges currently being proposed for power exports in South Australia (page 246 of \cite{AEMC2021}). 
We adopt this ratio of export to import charges ($2/13.2$) as fixed throughout the range of DUOS and LUOS tariffs in Fig.~\ref{combine}(b).

In both figures the gradients of the non-solar customers dark blue curves are lower than the solar customers light blue curves, indicating that non-solar customers are more strongly affected by LUOS tariffs. This is because non-solar customers import more of the solar energy that is exported into the local network than solar customers, whose solar generation profile is highly correlated to that of their neighbours.
Solar customers become more sensitive to the LUOS price - and the light blue curve becomes more horizontal - under the 2-way model because they are now charged for their exported solar.
This flattening of solar customers' light blue curve contributes to making the region of mutual benefit larger under the 2-way model than the 1-way model. 
The region is further increased by the CES more strongly mitigating solar customers' export charges by absorbing local solar exports, as seen by the light blue curve bending more strongly to the right as it crosses the LUOS~=~$0.5 \times$~DUOS line. 
The third factor influencing the larger region of mutual benefit is that the reference case for the 2-way model (marked by the purple square in Fig.~\ref{combine}(b)) exposes solar customers to additional networks fees on exports than the reference case of the 1-way model (purple square in Fig.~\ref{combine}(a)). 

Under the 2-way model, the average saving for solar customers is 3.0\% of their network fees when tariffs are set at: $\lambda_{u+}^n = 17$c/kWh, $\lambda_{l+}^n = 3$c/kWh, $\lambda_{u-}^n = 2.6$c/kWh, $\lambda_{l-}^n = 0.5$c/kWh. This saving is relative to current import only DUOS tariffs set at 13.2c/kWh. Their savings are even greater (9.3\%) when compared to SAPN's proposed DUOS only tariff settings of 13.2k/kWh on imports and 2c/kWh on exports \cite{SAPN}.
Given that non-solar customers aren't exposed to charges on exports their savings are significantly larger. They pay 22.2\% less in network fees under the 2-way LUOS model than under existing import only DUOS rates.
The savings under the 1-way LUOS model and the tariffs of Sect.~\ref{results1} are 1\% for solar customers and 17\% for non-solar customers, again relative to costs under current DUOS rates.

\subsubsection{Analysing factors at play for each stakeholder group}
\begin{figure}[]
\centering
\includegraphics[width=0.42\linewidth]{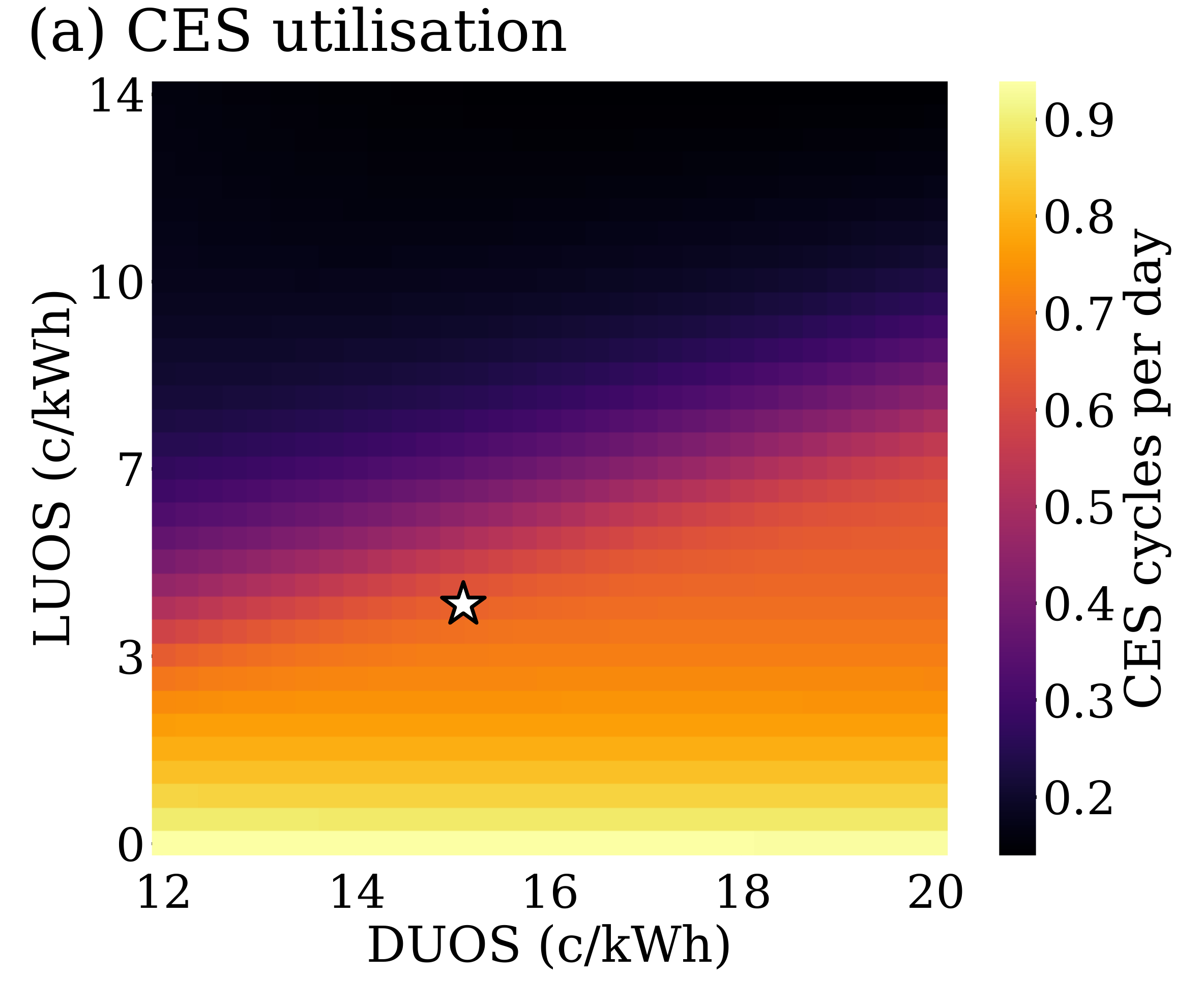}
\includegraphics[width=0.42\linewidth]{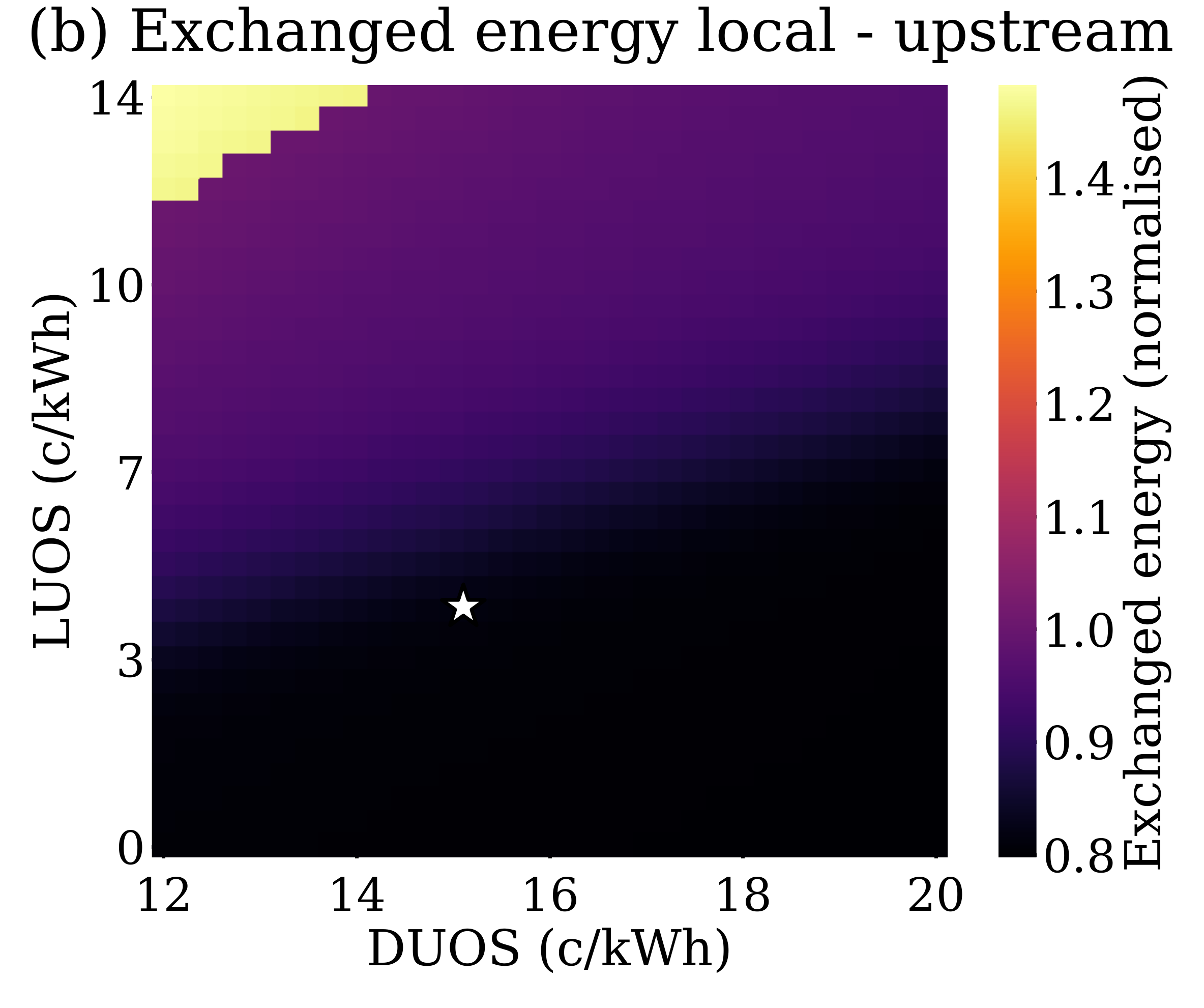}
\caption{Results for the 1-way LUOS model. (a) Utilisation of the CES as a function of DUOS and LUOS tariff values. Note the steep change as the LUOS takes on values less than half the DUOS (bottom right).
(b) Cumulative energy exchanged (imports and exports) between the local and upstream networks as a function of DUOS and LUOS tariffs, normalised to the reference case without a LUOS, where $\lambda_{u+}^n = \lambda_{l+}^n = 13.2$c/kWh, $\lambda_{l-}^n = \lambda_{u-}^n = 0$c/kWh.
The increased battery cycling in (a) is seen to clearly reduce the exchanged energy.
The top left corner of (b) shows how the perverse setting of LUOS~>~DUOS produces the perverse outcome of non-solar customers preferring to import energy from upstream rather than from their neighbours.
The white star in both figures indicates the tariff values studied in Sect.~\ref{results1} (DUOS = 15c/kWh and LUOS = 4c/kWh).
}
\label{battery_cycles}
\end{figure}

To understand how the conditions of mutual benefit arise we study the effects of DUOS and LUOS price settings on each stakeholder group.
Figure~\ref{battery_cycles}(a) examines the utilisation of the CES, calculated as the average number of CES cycles per day (charging and discharging the full storage capacity of the CES). It reveals that there is a striking increase in CES utilisation - from 0.15 times a day to over 0.9 times a day - when the LUOS rate becomes less that half the DUOS rate (the bottom right corner of the figure). 
This reflects the fact that all CES actions double the number of network charges applied cumulative to the community of customers and CES, as opposed to the comparative case where customers export their solar generation and later import energy from upstream (as detailed in Sect.~\ref{results0}).

\begin{figure}[]
\centering
\includegraphics[width=0.98\linewidth]{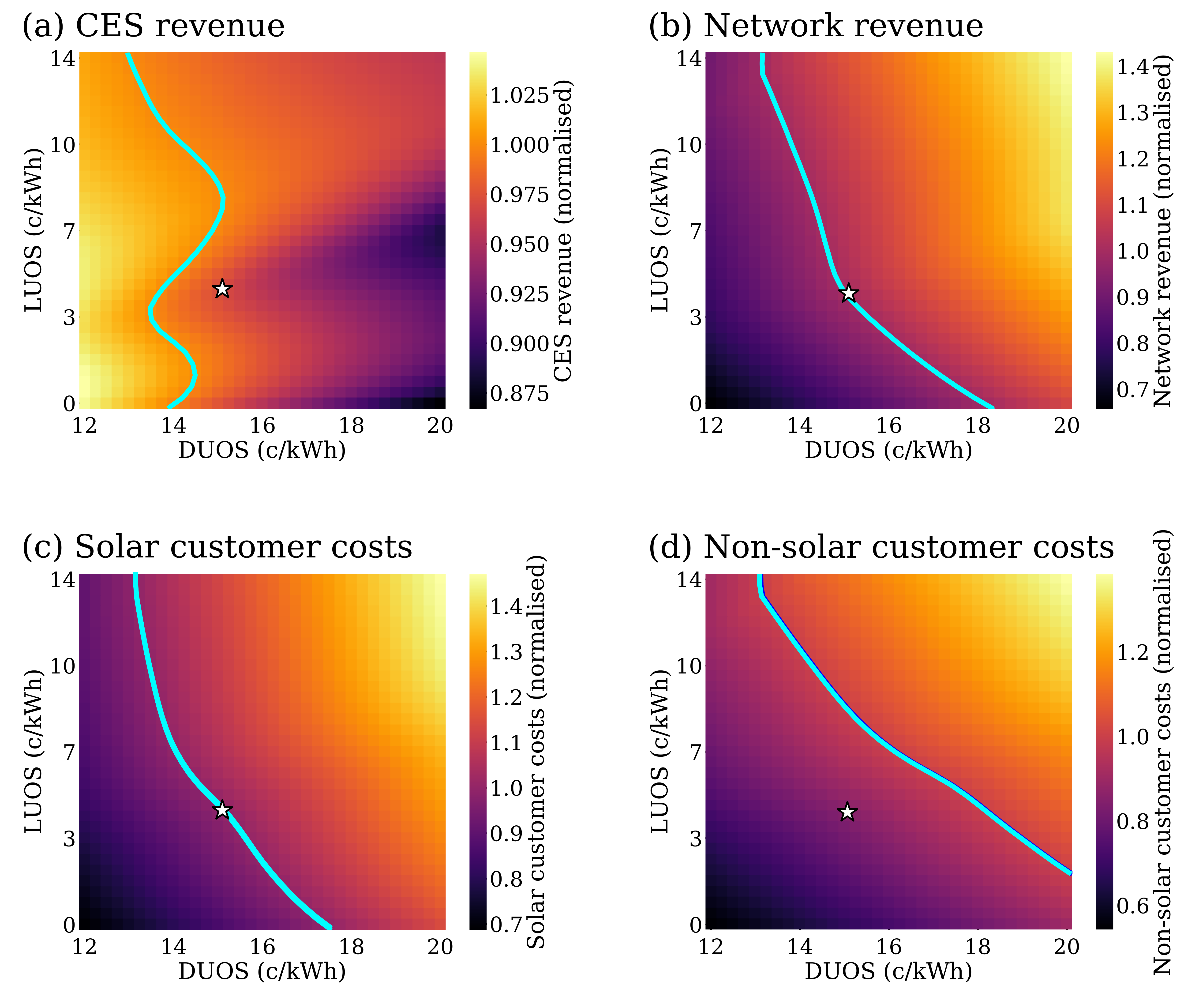}
\caption{Results for the 1-way LUOS model. (a) CES revenue and (b) distribution network operator revenue (c) 50 solar households customers and (d) 50 non-solar household customers as a function of DUOS and LUOS prices, normalised to the reference case without a LUOS, where $\lambda_{u+}^n = \lambda_{l+}^n = 13.2$c/kWh, $\lambda_{l-}^n = \lambda_{u-}^n = 0$c/kWh.
The cyan curves indicates tariff values that provide each stakeholder with equal revenue/costs to the reference case.
The white star indicates the tariff values studied in Sect.~\ref{results1} (DUOS = 15c/kWh and LUOS = 4c/kWh).
}
\label{customer_costs_broken}
\end{figure}

Mirroring the results for CES utilisation, Fig.~\ref{battery_cycles}(b) shows how the actions of the CES reduce the cumulative amount of energy exchanged between the local network and the upstream network ($E_{\rm ul} + E_{\rm ub} + E_{\rm gu} + E_{\rm bu}$). This confirms that discount LUOS tariffs incentivise the CES to absorb solar exports and displace upstream imports. Because the CES is operated to reduce the collective costs of customers and the CES, it does not always maximise it's own revenue. This can be seen in Fig.~\ref{customer_costs_broken}(a), which shows that the normalised revenue for the CES decreases at numerous tariff settings when LUOS~<~0.5~$\times$~DUOS and the CES prioritises reducing customers' network charges over arbitrage revenue.

\begin{figure}[]
\centering
\includegraphics[width=0.98\linewidth]{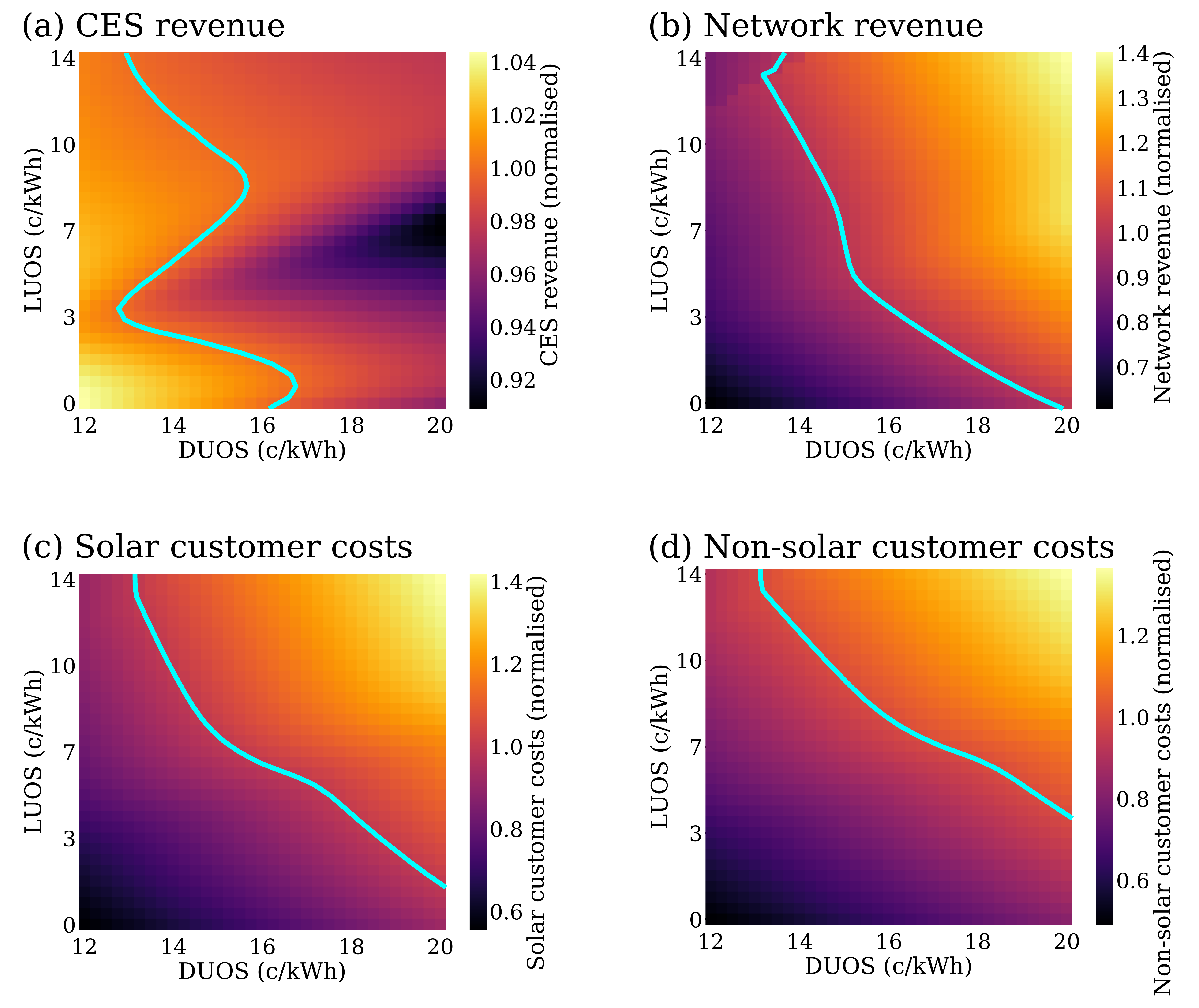}
\caption{Results for the 2-way model. (a) CES revenue and (b) distribution network operator revenue (c) 50 solar households customers and (d) 50 non-solar household customers as a function of DUOS and LUOS prices, normalised to the reference case without a LUOS, where $\lambda_{u+}^n = \lambda_{l+}^n = 13.2$c/kWh, $\lambda_{l-}^n = \lambda_{u-}^n = 2$c/kWh.
The cyan curves indicates tariff values that provide each stakeholder with equal revenue/costs to the reference case.
The white star indicates the tariff values studied in Sect.~\ref{results1} (DUOS = 15c/kWh and LUOS = 4c/kWh).
}
\label{customer_costs_broken2}
\end{figure}

Lastly, we present the results for the DN operator revenue and solar and non-solar customer costs in Fig.~\ref{customer_costs_broken}(b)-(d). Inspected individually, these highlight how the contour curves that represent constant costs/revenue to the reference case bend as they enter the region where LUOS~=~0.5~$\times$~DUOS (this bend is more prominent in the 2-way model discussed below). Crucially, the curves bend in opposite directions as they cross the LUOS~=~0.5~$\times$~DUOS line. This opens up the set of tariff values that produce savings for both groups of customers relative to the reference case (lying below the blue curves in Figs.~\ref{customer_costs_broken}(c),(d)), while producing revenues in excess of the reference case for the DN operator (lying above the green curve in Fig~\ref{customer_costs_broken}(b)).
This can be more clearly seen in the results for the 2-way model that are shown in Fig.~\ref{customer_costs_broken2}.

\clearpage
\subsection{Discussion}

In analysing the interactions of CES and LUOS our study extends these two fields. The demonstrated symbiotic interaction of the two innovations - under suitable tariff settings - makes their combined deployment more feasible, both technically and politically.
Our results demonstrate how LUOS tariffs improve the viability of CES and how the addition of CES into a network enables LUOS tariffs to create mutual benefits for all customers and for the DN operator.
Sections~\ref{results0} and \ref{results1} show how reducing network charges increases the utilisation of CES. Furthermore, reducing the charges through a LUOS tariff that is linked to the availability of local power exports and demand, shapes the additional utilisation to increase local self-sufficiency and reduce the load on the interconnection between the local and upstream networks.
The parameter sweeps of DUOS and LUOS tariffs in Sect.~\ref{results2} revealed that the additional energy transactions within the local network that occur when the CES charges and discharges to capture additional electricity market arbitrage value shift a significant portion of the cost of maintaining the DN onto the CES. This reduces the bills of customers (both those with and without solar), while maintaining the cost recovery of DN operators, who charge reduced per unit fees on an increased volume of energy flows.

We considered the simplified scenario where customers and the CES have direct access to a single, two-sided energy market, thereby eliminating the role of intermediary retailers. This is done in the interest of clarity, in terms of both the construction of retail offers and the relationship between customers of different retailers. 
It also more closely resembles existing market arrangements than P2P proposals that introduce a secondary energy market.
It also significantly eases the challenge of metering and discrimination between local and upstream energy flows. In this case a single meter at the point of coupling between the upstream and local network (together with existing metering at each customer and CES connection point) provides sufficient information to fully resolve the proportion of upstream and local energy flowing to each connection point. The involvement of multiple retailers (who do not share metering data) creates a substantial challenge to apportioning local/upstream energy flows to customers. These details do not negate the underlying value demonstrated in our study, but will add significant complexities in how this value may be realised by CES and customers. This will be explored in future work.

\clearpage
\section{Conclusion and Policy Implications}

The transition to a highly decentralised energy system involves many challenges, many of which consist of interwoven economic, technical and social issues.
This paper demonstrates how one such challenge - the management of two-way power flows in distribution networks - can be addressed through a symbiotic combination of an economic tool and the deployment of new energy assets. 
Specifically, local network tariffs improve the viability of CES by reducing their network tariff costs, while the CES transforms network reform from a zero-sum trade-off between customer segments and the network operator into an attractive proposition that can create mutual benefits for all stakeholders.
This is made possible by the CES accessing new arbitrage value in the energy market and, by creating additional transaction on the DN in the process, taking on a proportion of the costs of maintaining the DN.
This symbiosis resolves both the economic and social equity barriers facing each innovation on its own by being mutually beneficial to solar customers, non-solar customers \textit{and} the network operator. 
It provides policy makers with an economically and politically feasible solution for resolving the (otherwise fraught) issue of tariff reform for networks with large amounts of solar generation.

The clear recommendation from our mathematical analysis and simulations is that the price of local network tariffs (LUOS) needs to be less than half of conventional distribution network tariffs (DUOS). Only when this condition is met can the CES fulfil its potential to reduce customers' network charges by storing locally produced solar energy, as well as by arbitraging the energy market. Mutually beneficial economic outcomes for all stakeholders can only be achieved under this condition, and it is also a prerequisite for the CES reducing the strain on the distribution network.

Setting specific LUOS and DUOS tariff prices such that all stakeholders benefit will depend on the solar export and load characteristics of the local community and the size of the CES. This may make it challenging for regulators and DN operators to set LUOS and DUOS tariffs at appropriate levels throughout a DN in a way that is fair to all customers. Part of the solution to this may be to offer a range of tariffs, as is currently done with conventional network tariffs, although increased role of local conditions may make this more complex than for conventional tariffs.
Encouragingly, the smooth nature of the outcomes for stakeholders to variations in tariff prices (Figs.~\ref{battery_cycles}-\ref{customer_costs_broken2}) suggests that mutually beneficial settings are likely to be found relative to a wide range of reference tariffs.

The question of whether or not customers (and CES) ought to be charged for the power they export into the network is a political and regulatory choice that relates to issues of DER ownership as well as flexibility of shifting behaviour. Our results demonstrate that, were such tariffs to be adopted, the role of CES would be particularly impactful. Furthermore, we find that there are a significant range of 2-way tariffs that produce improved outcomes for all stakeholders relative to current import-only tariffs, including for solar customers.
The levying of network charges on exports would also be more pertinent where CES operated for profit maximisation. Under this goal the CES will only prioritise servicing local demand if it's network costs are lower for this action than for discharging power into the upstream network.
Network tariffs for exports may also be set so as to reward customers (and the CES) for exporting at times when this is likely to be beneficial to the system. The opportunities for such ToU tariffs with negative costs for export were identified by the AEMC \cite{AEMC2021}, although they have not yet been implemented by network operators. Incorporating such additional parameters of freedom will significantly expand the possibilities for networks to shape how DER interact with the network and will be the subject of future studies.

The scenarios considered in our study are unaffected by the ownership of the CES. They are however predicated on the ability of the CES to participate in the electricity market - as the arbitrage value from this market is what breaks the otherwise zero-sum distributions of costs amongst customers and the DN operator. Furthermore, the same logic implies that providing the CES with as many value streams as possible - such as participation in frequency control markets and network services markets - would further improve the outcomes for all stakeholders ans the CES generates increased revenue, increasing the volume of transaction on the local network and thereby decreasing the costs borne by customers. The use of LUOS network charges would bias the participation of the CES in all of these markets to respect the operating state of the local network.

Further studies are required to investigate the finer details of LUOS tariffs, such as the possibilities of ToU profiles, negative prices, and multiple tariffs for various customers. The politically charged issue of ownership of CES assets must also be resolved, or experimented with. The role of retailers and a heterogeneity of retail customers is also vital for mainstream deployments.

Overall, our study indicates that the joint deployment of CES and local network tariffs is a promising avenue for managing the two-way energy flows that arise in decentralised electricity systems. Deployed together, these innovations are technically, economically and socially appealing, effective and equitable.
Details of their deployments remain to be explored


\section*{Authorship Contribution Statement}
{\bf Bjorn Sturmberg:} Conceptualization, Methodology, Investigation, Writing - Original Draft; 
{\bf Marnie Shaw:} Conceptualization, Writing - Review \& Editing; 
{\bf Chathurika Mediwaththe:} Conceptualization, Writing - Review \& Editing; 
{\bf Hedda Ransan-Cooper:} Conceptualization, Writing - Review \& Editing; 
{\bf Benjamin Weise:} Conceptualization, Software, Writing - Review \& Editing;  
{\bf Michael Thomas:} Conceptualization, Software, Writing - Review \& Editing; 
{\bf Lachlan Blackhall:} Conceptualization, Methodology, Software, Writing - Review \& Editing.

\section*{Declaration of Competing Interest}
The authors declare that they have no known competing financial interests or personal relationships that could have appeared to influence the work reported in this paper.

\section*{Acknowledgement}
This work was supported by the Australian Renewable Energy Agency Advancing Renewables Program under grant 2018/ARP134.
We thank Reposit Power for providing the data used in this publication.

\appendix

\section{Appendix - optimization formulation}\label{appendix}
\newcommand{\soc}{\text{SoC}}
\newcommand{\capacity}{C}
\newcommand{\echg}{\eta^{+}}
\newcommand{\edchg}{\eta^{-}}
\newcommand{\throughput}{\$_\text{throughput}}
\newcommand{\egb}{E_{gb}}
\newcommand{\ebl}{E_{bl}}
\newcommand{\egl}{E_{gl}}
\newcommand{\ecl}{E_{ul}}
\newcommand{\ecb}{E_{ub}}
\newcommand{\ebc}{E_{bu}}
\newcommand{\egc}{E_{gu}}
\newcommand{\eimport}{E_{\text{net}}^{+}}
\newcommand{\eexport}{E_{\text{net}}^{-}}
\newcommand{\net}{E_{\text{net}}}
\newcommand{\smallm}{\epsilon}

In this paper, we are concerned with the optimization of real power ($P$) flows between various demand, generation and storage assets in a local network as given in Fig. \ref{framework}. The optimization will be performed in discrete time intervals where we assume that power is constant within each time interval. We denote an arbitrary time interval as $k$, whose duration is given by $\Delta k$. Without loss of generality we assume that $k \in \{1, 2, \dots, K \}$ and require that intervals are consecutive and non-overlapping so as to span a period of time completely. The energy transfer in the $k$-th time interval is given by $E = P \Delta k$.

\subsection{Constraints}

In implementing this optimization we impose the following constraints $\forall k \in \{1, 2, \dots, K \}$.

\subsubsection{Storage State of Charge}
The constraints in Eqs. \eqref{soc_efficiency} and \eqref{soc_limits} represent a simple energy storage model that capture both charging and discharging efficiency as well as ensuring that the energy storage capacity remains within allowable upper and lower bounds.
\begin{equation}
\soc(k) = \soc(k-1)+\echg E_b^+(k) + \frac{1}{\edchg} E_b^-(k)  
\label{soc_efficiency}
\end{equation}
where $\echg$ and $\edchg$ are the charge and discharge efficiency respectively, $E_b^+(k) = \egb(k) + \ecb(k)$, $E_b^-(k) = \ebl(k) + \ebc(k)$ and $\soc(0)$ is the initial state of charge. We also require that
\begin{equation}
    \underline{\soc} \leq \soc(k) \leq \overline{\soc}
    \label{soc_limits}
\end{equation}
where $\underline{\soc}$ and $\overline{\soc}$ are the minimum and maximum state of charge respectively.

\subsubsection{Storage Charge and Discharge Rates}
The constraint in Eq. \eqref{charge_limits} represents the charge and discharge limits of the energy storage device.
\begin{equation}
\begin{split}
    0 &\leq \frac{E_b^+(k)}{\Delta k} \leq \overline{P_b} \\
    \underline{P_b} &\leq \frac{E_b^-(k)}{\Delta k} \leq 0
    \end{split}
    \label{charge_limits}
\end{equation}
where $\underline{P_b}$ and $\overline{P_b}$ are the maximum discharge and charge rates, in kW, respectively.

\subsubsection{Local Network Constraints}
The constraints on the energy flows in the local network are defined graphically in Fig. \ref{framework} and supplemented by an additional constraint given in  \eqref{lem_local_flows} below.
\begin{equation}
\begin{split}
    \egl(k) + \egb(k) &\leq E_g(k) \\
    \egl(k) + \ebl(k) &\leq E_l(k)
    \end{split}
    \label{lem_local_flows}
\end{equation}
where $E_g(k)$ represents the net surplus generation from all customers that are exporting in the $k$-th time interval. Likewise, $E_l(k)$ represents the net unmet demand from all connection points that are importing energy in the $k$-th time interval.


\subsection{Objectives}
Having detailed the constraints used in the optimization we are now able to articulate the optimization objectives.

\subsubsection{Total Network Optimization}
We optimise to minimise the total cost to all participants within the local network which is defined in the following equation. The objective is minimised with respect to the decision variables which represent the energy flows to and from the battery and between surplus generation and demand.

\begin{equation}
\begin{split}
\min_{\ecb, \ebc, \egb, \ebl, \egl} \sum_{k=1}^K &(\lambda_{u+}^e(k) + \lambda_{u+}^n(k) )(\ecb(k) + \ecl(k)) \\
+~&(\lambda_{u-}^n(k) - \lambda_{u-}^e(k))(\ebc(k) + \egc(k)) \\
+~&(\lambda_{l-}^n(k) + \lambda_{l+}^n(k) - \lambda_{l-}^e(k) + \lambda_{l+}^e(k) ) \\
\times ~&(\egb(k) + \egl(k) +\ebl(k))
\end{split}
\label{lem_total}
\end{equation}

\subsubsection{Objective Penalty Terms}
We also describe two additional optimization objective terms that when added to the prior objective ensures more efficient and repeatable storage behaviours. The objective penalty detailed in \eqref{tput} adds a small penalty for energy flows into and out of the battery. This prevents unnecessary charge and discharge cycles from the energy storage that are uneconomic, and reflects that energy storage has a finite number of charge discharge cycles determined by the battery chemistry and warranty.
\begin{equation}
    \sum_{k=1}^K  (E_b^+(k) - E_b^-(k)) \times \frac{\throughput}{2}
    \label{tput}
\end{equation}
where $\throughput$ represents the round-trip throughput cost of storing energy.

We observe that the optimisation objective in \eqref{lem_total} converges to a solution set, not to a unique solution. To ensure a unique optimization solution, we introduce the penalty term in \eqref{storage_regularistion} which is a form of regularisation (as in \cite{ratnam2015optimization}) that will result in a unique optimization solution. With this additional penalty term, the objective becomes a quadratic program which can still be efficiently solved.

\begin{equation}
\begin{split}
    \sum_{k=1}^K  (&\ecb(k)^2 + \egb(k)^2 ~+ 
    \ebc(k)^2 + \ebl(k)^2) \times \smallm
    \end{split}
    \label{storage_regularistion}
\end{equation}
where the parameter $\smallm$ is chosen as a small number (i.e $1 \times 10^{-4}$).

\printcredits

\bibliographystyle{elsarticle-num}

\bibliography{main}

\begin{thebibliography}{10}
\expandafter\ifx\csname url\endcsname\relax
  \def\url#1{\texttt{#1}}\fi
\expandafter\ifx\csname urlprefix\endcsname\relax\def\urlprefix{URL }\fi
\expandafter\ifx\csname href\endcsname\relax
  \def\href#1#2{#2} \def\path#1{#1}\fi

\bibitem{CER2021}
{Clean Energy Regulator}, Postcode data for small-scale installations,
  \url{http://www.cleanenergyregulator.gov.au/RET/Forms-and-resources/Postcode-data-for-small-scale-installations}
  (2021).

\bibitem{AEMC2021}
{The Australian Energy Market Commission}, Draft rule determination national
  electricity amendment (access, pricing and incentive arrangements for
  distributed energy resources) rule 2021,
  \url{{https://www.aemc.gov.au/sites/default/files/2021-03/Draft%20Determination%20-%20ERC0311%20and%20RRC0039%20-%20Access%20Pricing%20and%20Incentive%20arrangements%20for%20DER.pdf}}
  (2021).

\bibitem{ZANDER2020111508}
K.~K. Zander, Unrealised opportunities for residential solar panels in
  australia, Energy Policy 142 (2020) 111508.
\newblock \href {https://doi.org/https://doi.org/10.1016/j.enpol.2020.111508}
  {\path{doi:https://doi.org/10.1016/j.enpol.2020.111508}}.

\bibitem{Cipcigan}
L.~Cipcigan, P.~Taylor, Investigation of the reverse power flow requirements of
  high penetrations of small-scale embedded generation, IET Renewable Power
  Generation 1 (2007) 160--166(6).
\newblock \href {https://doi.org/https://doi.org/10.1049/iet-rpg:20070011}
  {\path{doi:https://doi.org/10.1049/iet-rpg:20070011}}.

\bibitem{RUTOVITZ2018324}
J.~Rutovitz, S.~{Oliva H.}, L.~McIntosh, E.~Langham, S.~Teske, A.~Atherton,
  S.~Kelly, Local network credits and local electricity trading: Results of
  virtual trials and the policy implications, Energy Policy 120 (2018)
  324--334.
\newblock \href {https://doi.org/https://doi.org/10.1016/j.enpol.2018.05.026}
  {\path{doi:https://doi.org/10.1016/j.enpol.2018.05.026}}.

\bibitem{barbour_community_2018}
E.~Barbour, D.~Parra, Z.~Awwad, M.~C. Gonz{\'a}lez, Community energy storage: A
  smart choice for the smart grid?, Applied energy 212 (2018) 489--497.
\newblock \href {https://doi.org/10.1016/j.apenergy.2017.12.056}
  {\path{doi:10.1016/j.apenergy.2017.12.056}}.

\bibitem{Ransan-Cooper2021}
H.~Ransan-Cooper, B.~C.~P. Sturmberg, M.~E. Shaw, L.~Blackhall, Applying
  responsible algorithm design to neighbourhood-scale batteries in australia,
  Nature Energy 6 (2021) 815--823.
\newblock \href {https://doi.org/10.1038/s41560-021-00868-9}
  {\path{doi:10.1038/s41560-021-00868-9}}.

\bibitem{mediwaththe_network-aware_2021}
C.~P. Mediwaththe, L.~Blackhall, Network-aware demand-side management framework
  with a community energy storage system considering voltage constraints,
  {IEEE} Transactions on Power Systems 36~(2) (2021) 1229--1238.
\newblock \href {https://doi.org/10.1109/TPWRS.2020.3015218}
  {\path{doi:10.1109/TPWRS.2020.3015218}}.

\bibitem{SCHELLER}
F.~Scheller, R.~Burkhardt, R.~Schwarzeit, R.~McKenna, T.~Bruckner, Competition
  between simultaneous demand-side flexibility options: the case of community
  electricity storage systems, Applied Energy 269 (2020) 114969.
\newblock \href
  {https://doi.org/https://doi.org/10.1016/j.apenergy.2020.114969}
  {\path{doi:https://doi.org/10.1016/j.apenergy.2020.114969}}.

\bibitem{dong_impact_2020}
S.~Dong, E.~Kremers, M.~Brucoli, S.~Brown, R.~Rothman, Impact of household
  heterogeneity on community energy storage in the {UK}, Energy Reports 6
  (2020) 117--123.
\newblock \href {https://doi.org/10.1016/j.egyr.2020.03.005}
  {\path{doi:10.1016/j.egyr.2020.03.005}}.

\bibitem{Passey_CES}
R.~Passey, A.~Ngo, M.~Watt, J.~Jordan, J.~Zapata, {Business Models and
  Regulatory Considerations for Storage on the Distribution Network},
  \url{https://bit.ly/3gRWYwL} (2020).

\bibitem{langham2014}
E.~Langham, J.~Rutovitz, C.~Cooper, C.~Dunstan, Calculating the network value
  of local generation and consumption. report prepared for total environment
  centre and the city of sydney. (2014).

\bibitem{rutovitz2014level}
J.~Rutovitz, E.~Langham, J.~Downes, {A level playing field for local energy.
  Issues paper prepared for the City of Sydney.} (2014).

\bibitem{roy2016potential}
A.~Roy, A.~Bruce, I.~MacGill, {The potential value of peer-to-peer energy
  trading in the Australian national electricity market}, in: Asia-pacific
  solar research conference, 2016.

\bibitem{mengelkamp_role_2017}
E.~Mengelkamp, J.~Garttner, C.~Weinhardt, The role of energy storage in local
  energy markets, in: 2017 14th International Conference on the European Energy
  Market ({EEM}), 2017, pp. 1--6.
\newblock \href {https://doi.org/10.1109/EEM.2017.7981906}
  {\path{doi:10.1109/EEM.2017.7981906}}.

\bibitem{mengelkamp_trading_2017}
E.~Mengelkamp, P.~Staudt, J.~Garttner, C.~Weinhardt, Trading on local energy
  markets: A comparison of market designs and bidding strategies, in: 2017 14th
  International Conference on the European Energy Market ({EEM}), 2017, pp.
  1--6.
\newblock \href {https://doi.org/10.1109/EEM.2017.7981938}
  {\path{doi:10.1109/EEM.2017.7981938}}.

\bibitem{rodrigues_battery_2020}
D.~L. Rodrigues, X.~Ye, X.~Xia, B.~Zhu, Battery energy storage sizing
  optimisation for different ownership structures in a peer-to-peer energy
  sharing community, Applied Energy 262 (2020) 114498.
\newblock \href {https://doi.org/10.1016/j.apenergy.2020.114498}
  {\path{doi:10.1016/j.apenergy.2020.114498}}.

\bibitem{tushar_peer--peer_2021}
W.~Tushar, C.~Yuen, T.~K. Saha, T.~Morstyn, A.~C. Chapman, M.~J.~E. Alam,
  S.~Hanif, H.~V. Poor, Peer-to-peer energy systems for connected communities:
  A review of recent advances and emerging challenges, Applied Energy 282
  (2021) 116131.
\newblock \href {https://doi.org/10.1016/j.apenergy.2020.116131}
  {\path{doi:10.1016/j.apenergy.2020.116131}}.

\bibitem{guerrero_local_2019}
J.~Guerrero, A.~C. Chapman, G.~Verbič, Local energy markets in {LV} networks:
  Community based and decentralized p2p approaches, in: 2019 {IEEE} Milan
  {PowerTech}, 2019, pp. 1--6.
\newblock \href {https://doi.org/10.1109/PTC.2019.8810588}
  {\path{doi:10.1109/PTC.2019.8810588}}.

\bibitem{arghandeh_economic_2014}
R.~Arghandeh, J.~Woyak, A.~Onen, J.~Jung, R.~P. Broadwater, Economic optimal
  operation of community energy storage systems in competitive energy markets,
  Applied Energy 135 (2014) 71--80.
\newblock \href {https://doi.org/10.1016/j.apenergy.2014.08.066}
  {\path{doi:10.1016/j.apenergy.2014.08.066}}.

\bibitem{talent_optimal_2018}
O.~Talent, H.~Du, Optimal sizing and energy scheduling of photovoltaic-battery
  systems under different tariff structures, Renewable Energy 129 (2018)
  513--526.
\newblock \href {https://doi.org/10.1016/j.renene.2018.06.016}
  {\path{doi:10.1016/j.renene.2018.06.016}}.

\bibitem{van_der_stelt_techno-economic_2018}
S.~van~der Stelt, T.~{AlSkaif}, W.~van Sark, Techno-economic analysis of
  household and community energy storage for residential prosumers with smart
  appliances, Applied Energy 209 (2018) 266--276.
\newblock \href {https://doi.org/10.1016/j.apenergy.2017.10.096}
  {\path{doi:10.1016/j.apenergy.2017.10.096}}.

\bibitem{terlouw_multi-objective_2019}
T.~Terlouw, T.~{AlSkaif}, C.~Bauer, W.~van Sark, Multi-objective optimization
  of energy arbitrage in community energy storage systems using different
  battery technologies, Applied Energy 239 (2019) 356--372.
\newblock \href {https://doi.org/10.1016/j.apenergy.2019.01.227}
  {\path{doi:10.1016/j.apenergy.2019.01.227}}.

\bibitem{hafiz_energy_2019}
F.~Hafiz, A.~Rodrigo~de Queiroz, P.~Fajri, I.~Husain, Energy management and
  optimal storage sizing for a shared community: A multi-stage stochastic
  programming approach, Applied Energy 236 (2019) 42--54.
\newblock \href {https://doi.org/10.1016/j.apenergy.2018.11.080}
  {\path{doi:10.1016/j.apenergy.2018.11.080}}.

\bibitem{dong_techno-enviro-economic_2020}
S.~Dong, E.~Kremers, M.~Brucoli, R.~Rothman, S.~Brown, Techno-enviro-economic
  assessment of household and community energy storage in the {UK}, Energy
  Conversion and Management 205 (2020) 112330.
\newblock \href {https://doi.org/10.1016/j.enconman.2019.112330}
  {\path{doi:10.1016/j.enconman.2019.112330}}.

\bibitem{schram_trade-off_2020}
W.~L. Schram, T.~{AlSkaif}, I.~Lampropoulos, S.~Henein, W.~G. J. H. M.~v. Sark,
  On the trade-off between environmental and economic objectives in community
  energy storage operational optimization, {IEEE} Transactions on Sustainable
  Energy 11~(4) (2020) 2653--2661.
\newblock \href {https://doi.org/10.1109/TSTE.2020.2969292}
  {\path{doi:10.1109/TSTE.2020.2969292}}.

\bibitem{walker_analysis_2021}
A.~Walker, S.~Kwon, Analysis on impact of shared energy storage in residential
  community: Individual versus shared energy storage, Applied Energy 282 (2021)
  116172.
\newblock \href {https://doi.org/10.1016/j.apenergy.2020.116172}
  {\path{doi:10.1016/j.apenergy.2020.116172}}.

\bibitem{parra_optimum_2015}
D.~Parra, M.~Gillott, S.~A. Norman, G.~S. Walker, Optimum community energy
  storage system for {PV} energy time-shift, Applied Energy 137 (2015)
  576--587.
\newblock \href {https://doi.org/10.1016/j.apenergy.2014.08.060}
  {\path{doi:10.1016/j.apenergy.2014.08.060}}.

\bibitem{parra_optimum_2017}
D.~Parra, S.~A. Norman, G.~S. Walker, M.~Gillott, Optimum community energy
  storage for renewable energy and demand load management, Applied Energy 200
  (2017) 358--369.
\newblock \href {https://doi.org/10.1016/j.apenergy.2017.05.048}
  {\path{doi:10.1016/j.apenergy.2017.05.048}}.

\bibitem{pimm_community_2020}
A.~J. Pimm, J.~Palczewski, R.~Morris, T.~T. Cockerill, P.~G. Taylor, Community
  energy storage: A case study in the {UK} using a linear programming method,
  Energy Conversion and Management 205 (2020) 112388.
\newblock \href {https://doi.org/10.1016/j.enconman.2019.112388}
  {\path{doi:10.1016/j.enconman.2019.112388}}.

\bibitem{dong_improving_2020}
S.~Dong, E.~Kremers, M.~Brucoli, R.~Rothman, S.~Brown, Improving the
  feasibility of household and community energy storage: A
  techno-enviro-economic study for the {UK}, Renewable and Sustainable Energy
  Reviews 131 (2020) 110009.
\newblock \href {https://doi.org/10.1016/j.rser.2020.110009}
  {\path{doi:10.1016/j.rser.2020.110009}}.

\bibitem{Powerbank}
{Western Power}, Powerbank community battery storage,
  \url{https://westernpower.com.au/energy-solutions/projects-and-trials/powerbank-community-battery-storage/}
  (2019).

\bibitem{mediwaththe2019incentive}
C.~P. Mediwaththe, M.~Shaw, S.~Halgamuge, D.~Smith, P.~Scott, An
  incentive-compatible energy trading framework for neighborhood area networks
  with shared energy storage, IEEE Transactions on Sustainable Energy (2019).

\bibitem{mediwaththe2015dynamic}
C.~P. Mediwaththe, E.~R. Stephens, D.~B. Smith, A.~Mahanti, A dynamic game for
  electricity load management in neighborhood area networks, IEEE Transactions
  on Smart Grid 7~(3) (2015) 1329--1336.

\bibitem{mediwaththe2017competitive}
C.~P. Mediwaththe, E.~R. Stephens, D.~B. Smith, A.~Mahanti, Competitive energy
  trading framework for demand-side management in neighborhood area networks,
  IEEE Transactions on Smart Grid 9~(5) (2017) 4313--4322.

\bibitem{8513887}
J.~{Guerrero}, A.~C. {Chapman}, G.~{Verbič}, Decentralized p2p energy trading
  under network constraints in a low-voltage network, IEEE Transactions on
  Smart Grid 10~(5) (2019) 5163--5173.
\newblock \href {https://doi.org/10.1109/TSG.2018.2878445}
  {\path{doi:10.1109/TSG.2018.2878445}}.

\bibitem{CC2017}
{Climate Council}, Renewables ready: States leading the charge,
  \url{https://www.climatecouncil.org.au/uploads/9a3734e82574546679510bdc99d57847.pdf}
  (2017).

\bibitem{energymag}
{Energy Magazine}, Solar fuels south australia’s total energy demand in
  australian-first,
  \url{{https://www.energymagazine.com.au/solar-fuels-south-australias-total-energy-demand-in-australian-first/}}.

\bibitem{nemweb}
{Australian Energy Market Operator}, Market data nemweb,
  \url{https://www.aemo.com.au/energy-systems/electricity/national-electricity-market-nem/data-nem/market-data-nemweb}.

\bibitem{SAPN}
{South Australia Power Networks}, Tariff price list,
  \url{https://www.sapowernetworks.com.au/public/download.jsp?id=315323/}, the
  price list provides NUOS values that combine TUOS and DUOS. We set DUOS to
  80\% of NUOS, with the remainder being TUOS, based on the typical breakdown
  in Australia. The per kWh NUOS charge is the combination of the SUPPLY and
  ENERGY BASED USAGE values, dividing the former by the average daily household
  energy consumption of 19kWh. (2020).

\bibitem{Shaw2019}
M.~Shaw, B.~Sturmberg, L.~Guo, X.~Gao, E.~Ratnam, L.~Blackhall, The nextgen
  energy storage trial in the act, australia, in: Proceedings of the Tenth ACM
  International Conference on Future Energy Systems, Association for Computing
  Machinery, 2019, pp. 439--442.
\newblock \href {https://doi.org/10.1145/3307772.3331017}
  {\path{doi:10.1145/3307772.3331017}}.

\bibitem{Kapoor2020}
S.~Kapoor, B.~C.~P. Sturmberg, M.~E. Shaw, {A Review of Publicly Available
  Energy Data Sets},
  \url{https://wattwatchers.com.au/wp-content/uploads/2020/07/ANU_literature_Review_of_Energy_Data_Sets.pdf}
  (2020).

\bibitem{c3xgithub}
{The Battery Storage and Grid Integration Program}, c3x,
  \url{https://github.com/bsgip/c3x-enomo}.

\bibitem{throughput}
$\$_\text{throughput}$ = (capital cost per kWh $\times$ warrantied degradation)
  / (number of warrantied cycles). $\$_\text{throughput}$ = 3.2c/kWh when
  capital cost is A\$320 / kWh, warrantied degradation is 20\% and number of
  warranties cycles is 3650 (10 years).

\bibitem{ratnam2015optimization}
E.~L. Ratnam, S.~R. Weller, C.~M. Kellett, An optimization-based approach to
  scheduling residential battery storage with solar pv: Assessing customer
  benefit, Renewable Energy 75 (2015) 123--134.
\newblock \href {https://doi.org/https://doi.org/10.1016/j.renene.2014.09.008}
  {\path{doi:https://doi.org/10.1016/j.renene.2014.09.008}}.

\end{thebibliography}



\end{document}